\documentstyle[aaspp4,12pt]{article}
\def\Msun{\hbox{M$_{\odot}$}}
\def\Mbh{\hbox{M$_{\bullet}$}}
\def\Lsun{\hbox{L$_{\odot}$}}
\def\etal{\hbox{\it et al.}}
\def\imfa{\hbox{$\delta(m-1)$}}
\def\imfb{\hbox{$m^{-2.35}$}}
\def\rana{\hbox{1}}
\def\ranb{\hbox{[0.2,1]}}
\begin{document}

\title{The Dynamical Evolution of Dense Rotating Systems\\
Paper II.  Mergers and Stellar Evolution}
\author{John S. Arabadjis and Douglas O. Richstone}
\affil{University of Michigan Department of Astronomy}
\authoraddr{Ann Arbor, MI 48109}

\begin{abstract}

We report the results of simulations of dense rotating stellar systems
whose members suffer collisions and undergo stellar evolution processes.
The initial configuration for each experiment is an isotropic Kuzmin-Kutuzov
model.  The dynamical evolution is simulated with the N-body tree code of
Hernquist, modified to incorporate physical stellar collisions, stellar
evolution, stellar mass loss and compact remnant formation, and star
formation.  In some simulations we have added a large accreting
central black hole.  In all systems the velocity dispersion in the halo
evolves toward a radially biased state.  In systems containing a central
black hole, the dispersion becomes tangentially biased in the core, whereas
it remains isotropic in systems with no black hole.  Collisions tend to
produce a single dominant stellar merger product, as opposed to a swarm of
intermediate-mass stars.  In cases where we have suppressed all processes
except relaxation and physical collisions, objects with greater flattening
produce larger stars through mergers.  In systems where stellar ejecta are
allowed to escape the system, mass loss from the heavy core stars temporarily
reduces the core density and collision rate.

Most of the simulations performed reproduced the ratio of the central
collision time scale to the central relaxation time scale found in the dwarf
elliptical galaxy M32.  The rapid central evolution of these systems due to
collisions and relaxation, combined with scaling the results in N, suggests
that we are either viewing M32 at a peculiar moment in its history, or that
its dynamically-inferred central density is at least in part due to the
present of a massive dark object, presumably a black hole.

\end{abstract}


\section{Introduction}

The evolution of a dense stellar system is governed by two-body relaxation,
stellar evolution and mass loss, star formation, and stellar collisions.
In Paper I, we examined the effects wrought by two-body relaxation, and
the attendant mass segregation.  This study seeks to ascertain the relative
importance of the other physical processes in this list, as well as the
presence of a large black hole at the system center.  We simulated the
evolution using the N-body tree code of Hernquist (1987, 1990), modified to
include the processes listed above.

The motivation for this study is the dwarf elliptical galaxy M32, whose short
core relaxation and stellar collision time scales suggest the presence of a
central black hole (see Paper I for a discussion).  Recent observations and
dynamical modelling suggest a black hole mass of $3\times 10^6$ \Msun\
(van der Marel \etal\ 1997).  If M32 does not contain a black hole, it is
likely that a period of wholesale stellar merging and core collapse will
occur.  This would result in the formation of a large black hole (Quinlan and
Shapiro 1990) or a swarm of neutron stars, which could in turn collapse to
form a black hole (Quinlan and Shapiro 1989).  In the absence of significant
heating, the primary resistance to a core collapse episode is mass mass loss
from intermediate and high mass stars in the core.  Applegate (1986) has
shown, using a simple analytic model, that evolution of a cluster is very
sensitive to the slope of the initial mass function.  The results of
Chernoff and Weinberg (1990) indicate that the population of Galactic globular
clusters observed today may be only a subset of the primordial population, the
rest having evaporated as their heaviest members shed mass and the Galactic
tidal field stripped away their halos.  While a much more massive system like
M32 was probably never in danger of evaporating, we do expect its core
evolution to be sensitive to mass loss effects.

Once gas has been liberated by a star, its fate is not clear.  Whether the
material should be allowed to escape the cluster or should be reprocessed
into stars is an open question.  Observations of M32 show little or no gas
and dust (e.g. Tonry 1989), suggesting that supernovae and planetary nebulao
ejecta escape the system.  Supernova ejection velocities are generally far
in excess of the few hundred km s$^{-1}$ needed to escape from the core of
M32.  The presence of an interstellar medium undermines this argument,
however.  Galactic supernova remnants have been observed to sweep up 5 to 10
times their original mass (Strom 1988).  Conservation of momentum would
produce a corresponding reduction in speed, perhaps slowing core ejecta to
sub-escape velocities.  At expansion speeds of 10 to 100 km s$^{-1}$,
planetary nebula ejecta are even more likely to be confined to the galaxy.
In any case, the gas will not be swept up by the aggregate accretion of
member stars.  Because orbital speeds are greatly in excess of the sound
speed in any ambient medium, the Bondi accretion problem reduces to a
hard-sphere crossection (with a gravitational enhancement).  The instellar
medium would have to have a number density of $10^6$ in order for a star
to double its mass traveling at 100 km/s over $10^{10}$ years, far in excess
of that seen in M32.

Different studies have investigated a variety of models of the evolution of
the gas component.  Spitzer and Saslaw (1966) created solar neighborhood type
stars with their gas, while Sanders (1970) converted all ejecta into 0.5
\Msun\ stars.  Some have argued that material released in galactic nuclei
would form massive stars (Mathews 1972, Larson 1977), while others assume that
the ejecta to fall into a central black hole (David, Durisen, and Cohn 1987).
In perhaps the most complete Fokker-Planck (hereafter FP) study of dense
clusters to date, Quinlan and Shapiro (1990) made use of three scenarios for
the destination of their liberated gas -- complete system escape, reprocessing
into 1 \Msun\ stars, and reprocessing into stars following a Salpeter initial
mass spectrum (Salpeter 1955).  Because our code is strictly an N-body code,
with no algorithm for treating the complex gas dynamics involved in star
formation and stellar evolution processes, we have adopted two simple
scenarios:  in some simulations we have allowed the material to escape, and
in others we have converted it to Salpeter spectrum stars.

In addition to the processes discussed above, the presence of a large black
hole at the center of the system will profoundly affect the evolution of the
core.  Most of the work done on this problem has made use of analytic
calculations and FP methods.  Cohn and Kulsrud (1978) studied spherical
cluster evolution using single-mass component FP calculations.  Young (1980)
computed analytic models wherein a pre-existing black hole grows
adiabatically at the center of a spherical cluster.  Lee (1992) used FP
calculations to study the evolution of rotating clusters containing massive
black holes in their centers.  In this study we use N-body simulations to
investigate the core evolution.

\section{Initial simulation configuration}

As described in Paper I, each of our simulations begins with an isotropic
Kuzmin-Kutuzov model (hereafter KK) rotationally flattened to an ellipticity
$\epsilon \in [0,0.7]$ (Kuzmin and Kutuzov 1962; Dejonghe and De Zeeuw 1988).
In some cases we have modified the potential-density pair to include a central
black hole.  Defining \Mbh\ and M as the black hole mass and total mass,
respectively, the potential-density pair becomes

\begin{equation}
\phi(\varpi,z) = -\frac{G(M-\Mbh)}
{(\varpi^2+z^2+a^2+c^2+2\sqrt{a^2c^2+c^2\varpi^2+a^2z^2})^{1/2}}
- \frac{G\Mbh}{(\varpi^2+z^2)^{1/2}}
\end{equation}

\noindent and
 
\begin{eqnarray}
\rho(\varpi,z) & = & \frac{(M-\Mbh) c^2}{4\pi} \cdot \nonumber \\
& & \left[ \frac{ (a^2+c^2)\varpi^2 + 2a^2z^2 + 2a^2c^2 +
a^4 + 3a^2\sqrt{a^2c^2+c^2\varpi^2 + a^2z^2} }
{ (a^2c^2+c^2\varpi^2+a^2z^2)^{3/2} (\varpi^2 + z^2 + a^2 + c^2 +
2\sqrt{a^2c^2+c^2\varpi^2+a^2z^2})^{3/2} } \right] \nonumber \\
& & + \,\, \frac{\Mbh\delta(0)}{4\pi} \, .
\end{eqnarray}

\noindent Here $\varpi$, $z$, and $\phi$ are the usual cylindrical
coordinates, and $c$ and $a$ are scale lengths, with $c/a$ roughly
corresponding to the model's axial ratio (Dejonghe and De Zeeuw 1988).  As
discussed in Paper I, the particle velocities are obtained with the (modified)
KK density/potential pair and the Jeans equations under the assumption of
isotropy (Satoh 1980; Binney {\em et al.} 1990).  The initial stellar
coordinates are sampled according to the prescription laid out in Paper I.
The initial equilibrium of systems created in this way is demonstrated in
Figures \ref{cent_rel} and \ref{cent_rel_V}.  The former shows the core region
of a black hole model at $t/t_{rc}=$0, 1, 2, and 3, while the latter shows the
same but in velocity space.  The black hole is represented by a filled circle
near the origin.  There is little evolution over the first few local relaxation
times, indicating that the system is very much in equilibrium.

The mass spectrum used for each simulation is either a delta function or a
Salpeter spectrum.  For the simulations which use the Salpeter IMF we use
$m_{min}=0.2$ \Msun, rather than the standard 0.08 \Msun\ (e.g.
Grossman, Hays, and Graboske 1974), in order to populate the top decade of
the distribution for N=3000 (see discussion in Paper I).  We set
$m_{max}=100$ \Msun\ for some our models (Humphreys and Davidson 1986), but
for others we use 1 \Msun, since more massive stars would have perished by
now, given an age of $10^{10}$ y for M32.

\section{Physical Processes}

The physical processes which dominate the evolution of a dense stellar
system are two-body gravitational relaxation, stellar evolution and mass loss,
gas dynamics and star formation, stellar collisions, binary star interactions,
and external tidal fields.  In the case where the cluster contains a central
massive black hole, tidal disruption of low-$J$ stars will also affect the
cluster structure.

As discussed in Paper I, binary stars will not significantly alter the
evolution of the core of M32, and so they are not included in these
simulations.  Tidal fields may strip members from the outer halo of M32;
however, these simulations are geared primarily to investigate the core of
M32, and so we ignore tidal stripping.

\subsection{Two-body Relaxation}

The dynamical evolution of these systems was simulated using the N-body tree
code of Hernquist (1987, 1990), a {\sc Fortran} implementation of the
Barnes and Hut (1986) hierarchical force algorithm.  Two-body relaxation is
the scattering of stars off irregularities in the gravitational field of the
galaxy due to the finite number of stars.  The time scale associated with this
process is given by (e.g. Binney and Tremaine 1987, hereafter BT)

\begin{equation}
t_{\rm rc} = 0.337 \frac{\sigma^3}{G m_{\star} \rho \log\Lambda} \, .
\end{equation}

\noindent Here $\rho$ is the stellar density, $\sigma$ is the velocity
dispersion, $m_{\star}$ is the stellar mass, and $\log \Lambda$ is the
Coulomb logarithm.  Relaxation dominates the evolution of dense systems
whose stars have a range of masses (Spitzer and Saslaw 1966; H\'{e}non 1975;
Chernoff and Weinberg 1990; Giersz and Heggie 1996), but even for systems
with stars of equal mass relaxation will be extremely important because
stellar collisions will produce and subsequently flatten a mass spectrum.

\subsection{Stellar Evolution and Mass Loss}

Roughly 90\% of a star's life is spent on the main sequence (Iben 1964),
and so in all of our simulations we assume that a star will remain on the main
sequence until it sheds mass and becomes a compact remnant.  We adopt a power
law for the main sequence mass-radius relation:
 
\begin{equation}
\frac{r}{R_{\odot}} = \left( \frac{m}{\Msun} \right)^{\beta} \, .
\end{equation}
 
\noindent For low-mass ($0.1 < m < 2$ \Msun) main sequence stars, $\beta$ is
about 0.55 (Iben 1964; Bond, Arnett, and Carr 1984).  A stellar merger event
would probably produce a star which is temporarily bloated; however, since
the Kelvin-Helmoltz time scale (the time required for the bloated star to
settle down onto the main sequence) is much shorter than the collision time
scale in our simulations, we do not need to modify the mass-radius relation 
(see, e.g., Quinlan and Shapiro 1990).

We calculate the stellar lifetime from a parabolic fit to the data in
Iben (1964).  We adopt
 
\begin{equation}
(\tau-\tau_{0}) = \beta (\mu-\mu_{0})^{2}
\end{equation}
 
\noindent for $m \leq 100$ \Msun\ and $\tau = {\rm constant}$ for
$m \geq 100$ \Msun, where $\mu = \log_{10}(m/\Msun)$ and
$\tau = \log_{10}(t/10^{6}{\rm y})$.  Since the mass-to-light ratio approaches
a constant as $m$ increases, due to the Eddington limit, we require that

\begin{equation}
\lim_{m \rightarrow 100 \Msun} \frac{d\tau}{d\mu} = 0 \,\, .
\end{equation}
 
\noindent The parameters chosen were $\mu_{0}=2$, $\tau_{0}=0.5$, and
$\beta=0.87$.  For $m > 100$ \Msun\ we set $\tau=\tau_{0}$, which
corresponds to a lifetime of 3.16 My.  The Iben data and the parabolic fit
are shown in Figure \ref{lifetime}.

A star spends 90\% of its life on the main sequence, or near enough to it that
its radius has not changed appreciably (Iben 1964, 1967).  For this reason,
and because giants have extremely tenuous atmospheres, stars in all of our
models are considered to be on the main sequence until a mass-loss event at
their death, for purposes of calculating their collision cross section.
Reimers and Koester (1982) have argued that stars of $m<8$ \Msun\ will leave
behind white dwarfs as their stellar evolution remnant, although stars in the
upper part of this range could explode as carbon-detonation supernovae, leaving
no remnant (see Iben and Renzini 1983 for review).  We choose to assign white
dwarf remnants to these stars.  Since the white dwarf mass distribution is
strongly peaked at 0.6 \Msun\ (Weidemann 1990), we use this value as our
remnant mass.  The rest of the stellar material is either ejected from the
cluster on a very short time scale, or it is reprocessed into stars, depending
on the model.  The stars created from ejecta are assigned coordinates according
to the initial $(m,{\bf x},{\bf v})$ distribution.

Bethe (1986) has argued that stars with masses $8\leq m< 25$ \Msun\ die in
supernova explosions, leaving neutron star remnants, while stars of mass
$m>25$ \Msun\ produce black hole remnants.  We adopt this view (because it
is simple) and set all neutron star masses to 1.4 \Msun, consistent with
binary pulsar measurements (Hulse and Taylor 1975, Taylor and Dewey 1988,
Taylor and Weisberg 1989).  The theoretical upper limit for the neutron
star mass is about 3 \Msun; it is somewhat less when more elastic (realistic)
equations of state are assumed (Chitre and Hartle 1976).  Several observed
binary systems show companions whose masses exceed this limit, suggesting the
presence of a black hole (for a review see Cowley 1992).  The four most
robust stellar-mass black hole candidates are listed in Table \ref{bhcands}.
In our simulations we set the remnant black hole mass equal to $0.2m$ \Msun,
producing holes with masses $\geq 5$ \Msun, with the remaining mass ejected as
a supernova explosion (Bodenheimer and Woosley 1983; Woosley and Weaver 1986).

\subsection{Star Formation}

As mentioned previously, our simulations utilized two procedures for treating
mass ejecta.  For the simulations in which ejected material is converted into
stars, the stellar masses and positions were assigned according to the
prescription described in Paper I.  At the end of each time step during which
there was at least one mass loss event, the ejecta are pooled together.  After
as many stars on the appropriate mass interval have been formed, the remaining
material is assigned to a single star on the interval (0,0.2] \Msun.

\subsection{Stellar Collisions}

The mean time for a star of radius $r_{\star}$ and mass $m$ to suffer a
physical collision with another star in a stellar system with number density
$n$ and dispersion $\sigma$ is given by (BT)

\begin{equation}
\label{eq:coll}
t_{\rm coll} = \left[ 16 \sqrt{\pi} n \sigma r_{\star}^{2}
(1+\Theta) \right]^{-1} \, .
\end{equation}

\noindent Recent observations (van der Marel \etal\ 1997) show that the central
velocity dispersion of M32 $\sigma_0 = 126$ km s$^{-1}$.  A stellar
collision in this range would most likely result in the merging of the two
stars, with little mass loss, for a wide range of stellar masses (Benz and
Hills 1987, 1992; Lai, Rasio, and Shapiro 1993).  Thus we assume that stellar
collisions in the core of M32 result in merger events, with no mass loss.

Figure \ref{tcol_local} shows collision and relaxation time scale contours for
a wide range of velocity dispersion and density.  The core properties of a
few stellar systems are also shown.  The solid symbols assume that the core
consists of main sequence stars only -- there would be a moderate collision
rate enhancement for cores containing a significant fraction of binary stars.
The open symbol for M87 uses the observed central star light density (5000
\Lsun\ pc$^{-3}$); the closed symbol assumes that the inferred mass density
is due solely to the presence of 1 \Msun\ stars.  (This is clearly not the
case -- it has been known for decades that the nucleus of M87 is quite active.)
The open symbol for M32 evaluates the cusp model of Lauer \etal\ (1992) at
$r = 0.1$ pc and the solid symbol represents the 0.37 pc core model of the
same study.

A study of the effects of stellar collisions on dynamical evolution would
naturally focus on those systems found to the right of the $t_{col}$ = 10 Gy
contour.  As this figure suggests, globular clusters generally do not possess
the core properties required for stellar collisions to be important.  The
exception to this is NGC 6256, whose high central density and low dispersion
make it an interesting laboratory for studying the effects of collisions.
Globular cluster evolution can be strongly influenced by the presence of
binary stars, however.  Since the tree code is not equipped to handle binaries
we have restricted our study to M32.

\subsection{Central black hole accretion}

Collisions of stars with a central black hole are treated in a different
fashion.  We assume that stars which venture too close to the hole are
disrupted by tidal forces.  If ${\bf x}_{\star}$ and ${\bf x}_{\bullet}$ are
the star and black hole positions, respectively, then tidal disruption occurs
if
 
\begin{equation}
\label{eq:roche}
|{\bf x}_{\star}-{\bf x}_{\bullet}| \, \le \, r_{\star}
\left( \frac{3\Mbh}{m_{\star}} \right)^{1/3} \, .
\end{equation}

\noindent Thus there are three relevant mass parameters for the system:
\Mbh/M$_{total}$, m$_{\star}$/\Mbh, and m$_{\star}$/M$_{total}$.  Due to the
necessity of simulating M32 with far fewer stars than it actually contains, we
cannot preserve all three of these ratios simultaneously.  At this point it is
not feasible to simulate M32 with an N-body code using N=$10^8$, so we seek
only to preserve the first two of these ratios.  First, we seek to simulate
only the inner 10\% of the system so that M$_{total}$ is a tenth of the actual
system mass.  Using M32 as our guide (e.g. van der Marel 1997), we set
\Mbh/M$_{total}$=0.1.  The \Mbh\ in equation \ref{eq:roche} is in essence a
free parameter, unrelated to \Mbh/M$_{total}$.  Thus we set
\Mbh/m$_{\star}=10^6$ in order to characterize the tidal dispruption and
black hole accretion properly.

\section{The Models}

The effects of stellar collisions on the dynamical evolution of dense systems
are quite tangled.  On the one hand, collisions will tend to produce a
high density core.  Heavy stars formed through mergers will sink to the
center of the system as they attempt to come to energy equipartition with
the surroundings.  In addition, the completely inelastic merging of a pair
of unbound stars removes energy from the system, which in turn deepens the
central potential (and mass density).  For a monotonically increasing
mass-radius relation for main sequence stars, as orbits of heavy stars decay,
their increasing cross section and density will enhance the frequency of
collisions.  In the absence of other processes a single high mass star may
form, merging with all others which migrate to the core.

On the other hand, collisions in the core can help indirectly to terminate
the infall of heavy stars.  High mass stars are short-lived, and if the
collision time scale exceeds the stellar evolution time scale, mass loss
tends to remove material from the core, even if it does not escape the system
as a whole.  This makes the central potential more shallow, and the core, no
longer in virial equilibrium, begins to expand.

In order to discriminate among these effects we sought first to understand
the influence of stellar collisions on system evolution, in the absence of
mass loss.  In particular, we wanted to know if the collision process favored
the formation of a single stellar merger remnant or a swarm of intermediate
mass objects, for a ratio of relaxation and collision time scales found in
astronomical objects.  We designed the input models such that
$t_{col}/t_{rel} \simeq 100$, as in the case in the core of M32.  We also
wanted to know if the flatness of the system influences the collision rate
or the mass of the merger remnants.  We performed eight simulations of 300
bodies each (Group 1).  The input models varied in initial flatness from
$c/a=1.0$ to 0.3, roughly corresponding to axisymmetric elliptical galaxies
E0 through E7, respectively.

The next set of experiments we ran (Group 2) were designed to explore
the effect of mass loss and stellar collisions on system evolution, as
well as the influence of a central black hole.  We decided to run four
simulations of 3000 bodies each, two with a central black hole (using
\Mbh=0.1M, similar to a possible nuclear configuration of M32) and
two without black holes.  Since we were interested in rotation effects,
we ran each with $c/a=1.0$ and 0.3.  We opted to use equal-mass stars in
the input models to reduce the noise associated with a mass spectrum.  This
also allows us to compare the black hole results with analytic models in
the literature.

The last group of experiments we performed (Group 3) were specifically
designed to mimic as many of the characteristics of M32 that our dynamic
range would allow.  We set $c/a=0.8$, $t_{col}/t_{rel}=100$, and we used
a Salpeter spectrum for the initial mass function.  We looked at the
effect of star formation (recycling of stellar ejecta) on the core
structure and the N dependence of the results.

Table \ref{modpars} summarizes the model parameters for each of the
simulations.  In the second column $a$ and $c$ are the relevant scale lengths
in the KK model.  Column 3 lists the initial number of bodies in each
simulation, and column 4 indicates the mass spectrum used.  The stellar mass
range is given in column 5, and columns 6-8 indicate which astrophysical
processes (stellar evolution, physical stellar collisions, and star formation,
respectively) were incorporated in each simulation.  The ratio of the collision
time scale to the relaxation time scale for the core is given in column 9,
and the central black hole mass is listed in column 10.

The last model listed under Group 3 (model XVIII) was not designed to
reproduce the probable ratio of the collision to relaxation time scales in
M32.  Rather, the simulation was performed to examine the types of stellar
remnants produced for a value of $t_{col}/t_{rel}$ which, although it is
unlikely value for the unresolved core of M32, is still a plausible value
for some dense stellar systems.

\subsection{Stellar Collisions and Merger Remnants}

We first examined the collision process in absence of all other physical
processes except relaxation.  We ran eight simulations (Group 1) for 100
half-mass relaxation times, with $t_{ch}/t_{rh}=100$.  We varied $c/a$
along the sequence, ranging from 1.0 through 0.3 (corresponding to
E0 through E7).

Because $t_{rh} \ll t_{ch}$ and $t_{rh} \ll t_{evo}$ ($t_{evo} \equiv \infty$
for this set of experiments), each merger remnant will fall to the center
before it collides again.  Therefore when they do collide a second time they
are more likely to collide with another merger remnant, since the core of the
system will be populated with the heaviest stars.  This effect is shown in
Figure \ref{treecol}, which shows the location of collision events in the
spherical case (model XVIa) versus time.  Events marked with an `x' denote
collisions between 1 \Msun\ stars, and triangles represent collisions between
stars whose masses sum to 10 \Msun\ or less (but greater than 2 \Msun).
Circles denote collisions between stars whose mass sum is greater than 10
\Msun; this last category amounts to collisions between field stars and the
dominant merger remnant.  Open circles represent collisions of 1 \Msun\ stars
with the merger remnant, and filled circles are collisions between the dominant
remnant and heavier stars.  Collisions between 1 \Msun\ stars occur over a much
wider range than collisions between heavier stars.  Collisions between
intermediate mass stars occur during a short phase near $t=100$, after which
the dominant merger remnant is a participant in most of the collision events.
The collision evolution in the more flattened systems follows the same
patttern.

The mass of the merger remnant appears to be mildly dependent upon the
rotational state, with flatter systems producing somewhat larger dominant
merger remnant.  The mass of the next largest (secondary) star appears to be
independent of the flattening.  The merger remnant and secondary star masses
at simulation termination are shown in Figure \ref{mmaxeps_cor}.

The reason for the the rotational dependence of the collision process can be
traced to the early evolution of the core.  It is well known that if the
characteristic rotation speed is too large, the system is unstable to $l$=2,
$m$=2 modes (ellipsoidal ``bar'' structures).  Kalnajs (1972) and Ostriker and
Peebles (1973) showed that the criterion for stability against the formation
of bar modes can be written

\begin{equation}
\frac{T}{|W|} < t_{crit} \, ,
\end{equation}

\noindent Here $T$ is the kinetic energy associated with the rotation and $W$
is the potential energy.  Kalnajs (1972) showed analytically that, for a
certain class of disk models which bear his name, $t_{crit}=125/486=0.1286$.
Ostriker and Peebles (1973) used N-body simulations of Mestel disks to derive
$t_{crit} \simeq 0.14$.  $T/W\simeq 0.16$ at $t=0$ in our E7 KK model (XVIh),
indicating that it is bar-mode unstable.  In Figure \ref{TW} we show the
evolution of $T/|W|$ for the this model.  The system quickly forms a bar
structure which transports angular momentum outward, allowing the core to
contract more rapidly and driving up the central density.  The bar cannot be
seen simply by viewing the system from above, due to the low number of bodies.
It can be seen, however, by comparing the mass of stars in each quadrant
(I--IV) as the cartesian axes are rotated by an amount $\delta\phi$.  If the
system is indeed centered on the origin, and a true bar structure exists, then
the quantity $(m_{I}+m_{III}/m_{II}+m_{IV})$ should show be sinusoidal in the
variable $\delta\phi$ (Fig. \ref{bar3schem}).  Figure \ref{bar3} shows this
quantity for model XVIh, clearly demonstrating the existence of a bar.

The increase in the density enhances the collision rate, which in turn builds
up the dominant merger remnant.  It is impossible to distinguish between the
two processes -- core contraction and stellar mergers -- since they enhance
each other.  Their influence can be observed in their effect on the density
profile.  The density of the two extreme cases, $c/a=1.0,0.3$, is shown at
three different times in Figure \ref{deps}.  The dominant merger remnant is
found within the innermost data point in each plot.

The increased central density allows the flatter rotating model to experience
its collision phase earlier in its evolution than a more spherical model.
Figure \ref{coll_ip} shows the collision rate for the two extreme cases.
Both systems experience an episode of enhanced collision activity; that of
the flattened system comes at a slightly earlier time, however, because of
the more rapid increase in the core density.  Relaxation causes the core of
a system to shrink in mass as well as size; since collisions occur in the
core, the earlier the episode occurs, the more stars available for merging.

Under the conditions explored in these simulations, with $t_{col}/t_{rel}=100$,
only one experiment produced a compact remnant other than a white dwarf.
(Model XII produced a neutron star in the vicinity of the central
black hole.)  This is not due to the physical conditions present, although it
is true that if the Salpeter systems with masses on the range [0.2,100] \Msun\
were allowed to undergo stellar evolution processes, we would likely produce
a black hole or two and many neutron stars, since there would be many stars
with the requisite initial mass.  The failure of these simultions to produce
neutron stars and black holes is simply due to the limited dynamic range of
the simulations.  Because of time constraints, we were forced into the
trade-off between the number of stars in a simulation and the number of
simulations.  We settled on a low N so that we could explore a range of
parameters relevant to dense elliptical galaxies.  One effect of low N is
to directly limit the mass of the dominant merger remnant.  Obviouly a system
of mass $M$ cannot form a remnant with a mass exceeding $M$.  If a system is
allowed to evolve for a period equal to its half-mass collision time scale,
one might expect that the merger remnant mass $m_{rem}$ would be some fraction
of the total number of stars available to the collision process, N/2.  We
therefore write

\begin{equation}
m_{rem} = \epsilon \, \frac{N}{2} \, ,
\end{equation}

\noindent where $\epsilon$ is an efficiency which depends on at least three
quantities -- the stellar evolution, relaxation, and collision time scales.
From our Group 2 simulations we can estimate $\epsilon$ in the case where the
stellar evolution time scale is infinite.  The largest stellar mass to form
was made from 29\% of the $N/2$ particles nominally available for merging.
When stellar evolution effects are included, the efficiency suffers
dramatically.

The dominant merger remnant produced in some of the simulations is listed in
Table \ref{domstarmass}.  In Figure \ref{mergeeff} we plot the merger
efficiency versus the number of stars in the system for XIII and XV.  The
points appear to follow a power law with a slope of about -0.7.  In the
absence of any new astrophysical processes becoming important on the interval
$3\times 10^3 < N < 10^8$, two interesting mass scales become apparent.  In
order for a system to produce a black hole through stellar collisions, the
quantity $\epsilon$N$/2$ must equal or exceed 25\Msun.  The minimum N
required to do this is is about $2\times 10^6$.  This value exceeds most
globular cluster masses, suggesting that globular clusters are not able to
form black holes in their cores through stellar mergers.  For a system
with N=$3\times 10^8$, a star of mass 100 \Msun\ can be produced through
collisions, consistent with the results of Quinlan and Shapiro (1989).

As an alternative to increasing N (or scaling N through a power law fit)
to produce a central black hole through stellar collisions is to reduce
$t_{ch}/t_{rh}$.  If we have an unresolved galaxy core like M32
(Lauer \etal\ 1992), it is possible (though unlikely -- see, e.g., Goodman
and Lee 1989) that the the true central number density is much larger than the
lower limit we are able to assign.  In that case, if we scale up the density,
we scale up the age of the galaxy, measured in units of the collision time
scale.  Unfortunately, the relaxation time scales with density roughly as does
the collision time; the ratio does not change.  However, in the regime of
non-disruptive stellar collisions,
 
\begin{equation}
t_{ch}/t_{rh} \propto \sigma^{-2} \, .
\end{equation}

\noindent If the central dispersion of M32 were a factor of 3 higher than
current observations suggest (again unlikely), then we could reduce the ratio
to 10 from 100.  While this is probably not realistic for M32, it may be for
other systems.  With this in mind, we ran an additional simulation (model
XVIII) which set $t_{ch}/t_{rh} = 10$, but which in all other respects
conformed to the parameters of model XV.

Between $t = 0$ and $100 t_{dh}$ the core collision rate is about 4
collisions per $t_{dh}$.  At $t=100t_{dh}$, two black holes are formed
($m$=10.6, 6.2 \Msun); accompanying this is the shedding of 67.2 \Msun\ of
material, which is 6\% of the total mass.  Although this material is
reprocessed into stars, it is lost, at least temporarily, from the core,
dramatically reducing the core density and collision rate.  In addition to
the pair of black holes, the system produced nine neutron stars, four white
dwarfs, and an additional 553 main sequence stars from recycled ejecta.  The
merger efficiency for this system was about 5\%.  While the simulation
may have little relevance to M32, it is nonetheless interesting that large
objects can be produced through merger events with only a moderate, and
plausible, adjustment to the model parameters, even within a low N system.

\subsection{Central Black Hole Models}

Next we contrast the evolution of systems which do or do not contain black
holes in their nuclei.  We expect that the density profiles of the core of
each system would show evidence of a massive object within the black hole
``cusp radius'', defined by

\begin{equation}
r_{cusp} = \frac{G \Mbh}{\sigma^2} \, ,
\end{equation}

\noindent Cohn and Kulsrud (1978) used Fokker-Planck calculations of spherical
clusters to derive an $r^{-7/4}$ density profile.  Young (1980) modeled a
spherical cluster containing a central black hole under the assumption that
the black hole grows adiabatically.  He found that the density profile goes
as $r^{-3/2}$ for a completely isotropic core and $r^{-9/4}$ for a core whose
stars all lie on circular orbits.  More recently, Quinlan \etal\ (1995)
calculated models for black holes in spherical clusters, and derived similar
results.  They found that $\rho \sim r^{-3/2}$ near the black hole in a model
whose core initially resembles an isothermal sphere.  The system retained an
isotropic velocity dispersion near the black hole; a tangential bias was seen
in the dispersion away from the hole.  For their non-isothermal models, they
found that the density cusp was steeper, and that the anisotropy penetrated
right into the center.

In Figure \ref{den_deBH} we show the density profiles for our simulations
which contain a central black hole and equal-mass stars.  We fitted a power
law of slope $\alpha$ to the inner 15\% of each system (neglecting the
innermost point containing the black hole), which falls within the black
hole cusp radius.  The data points used in the power law fit are shown as
circles or triangles in each graph.  (In each case $r_{cusp} \simeq 2.5$.)
For the spherical system (model XI) we found that $\alpha=-1.79\pm0.08$; the
flattened system (XII) had a slightly steeper profile at $\alpha=-2.15\pm0.06$.
The initial and final profiles of model XI are shown in Figure \ref{den_d_evo}
to demonstrate the evolution from a core-like to a cusp-like profile.

Next we show the density profile for four models with a Salpeter IMF and
rotationally flattened to $c/a=0.8$.  The first two (models XIIIa and XVII,
Figure \ref{den_qf}) contain no central black hole.  Model XVII, which
evolves under the influence of two-body relaxation only, has a somewhat more
shallow core profile than does model XIIIa, which incorporates stellar
evolution and collisions.  This is due to the fact that stellar collisions
produce a flatter effective mass function, and the subsequent mass
segregation moves heavier stars into and lighter stars out of the core,
increasing the density gradient.  The simulations of Figure \ref{den_ggBH}
each contain a black hole of mass \Mbh/\Msun=0.1.  Model XIVa, which
incorporates stellar evolution (mass loss) and merging, has a steeper core
profile, again due to the effects of mass segregation.  In simulations which
evolve under the influence of relaxation only (XVII and XIVb), the system
which contains a black hole (XIVb) develops a steeper core profile.  As in
the case without central black holes, the system which allows stellar
collisions and evolution produces a steeper profile than the system which
does not.

Models with and without central black holes show a marked difference in
the behavior of their velocity dispersions.  Each of the simulations
started with isotropic dispersions; relaxation and mergers drive the systems
away from isotropy.
 
We plot the dispersion ratios $\sigma_{\theta}/\sigma_{r}$ and
$\sigma_{\phi}/\sigma_{r}$ for each Lagrangian quintile of the models
which contain no central black hole, in Figure \ref{sigrt_ac}.  (The
curves have been smoothed over three bins.)  These systems tend to preserve
an isotropic dispersion in their cores, while their halo dispersions
become radially biased outside the half-mass radius.

The spherical black hole model (XI), on the other hand, develops a strong
tangential bias.  In Figure \ref{sigrt_de} we do not plot the innermost
quintile because it is quite noisy due to the fact that this mass quintile
initially houses the black hole, and eventually {\it becomes} the black hole
when the hole mass reaches 0.2 through accretion.  In the rotating black
hole model ($c/a=0.3$, model XII), the core of the system appears to remain
isotropic.  The reason for this is unclear.  It is not related to the
similar finding of Quinlan \etal\ -- their system behaved in such a way due
to an isothermal initial core structure, in contrast to their
non-isothermal core model.  In our systems, however, the cores start off
identically; they both possess isotropic cores, with only the flattening
differing between the two.  The plummeting dispersion ratio of the outermost
Lagrangian quintile is due to the presence of stars which have escaped the
system.

\subsection{M32 Simulations}

The simulations of Group 3 were specifically designed to mimic the evolution
of M32.  Assuming that M32 is 10 Gy old, we simulated it with a Salpeter
spectrum with $m_{max}=1.0$ \Msun, since presumably stars with masses greater
than 1 \Msun (neglecting blue stragglers formed through collisions) would
have perished by now.  (We note, however, that there is some evidence for the
the presence of a 5 Gy old population -- see Freedman 1989.) We scaled the
stellar radii to produce a time scale ratio $t_{col}/t_{rel}=100$.

An important issue when simulating M32 is to understand how the decision to
treat mass loss from the stars affects the outcome of experiments.  As
it turns out, the recycling of stellar ejecta into new stars does not have
a straightforward effect on the evolution of the core.  One might expect that
star formation would simply prevent the core from evaporating as heavy stars
therein shed mass.  The picture is somewhat more complicated.

Initially, mass recycling does prolong the episode of core contraction
(Figure \ref{core_r_recy}).  The core density of a recycling model
continues to rise past the point where the non-recycling system begins to
expand.  However, the increased amount of material in the core results in a
larger average stellar mass in the core.  These larger stars lose
proportionately more mass when they become white dwarfs, and the net effect
is that the core density will be {\it lower} when mass ejecta are recycled
into stars, for a period of time equal to that during which it was higher
(Figure \ref{core_p_recy}).  After 100 relaxation times, or about 2800
dynamical times, the recycling system density again exceeds the
non-recycling density.  This corresponds to about $10^{10}$ years,
indicating that the present core density of an elliptical galaxy cannot be
used to determine whether mass loss from stars in the core escapes the
system.

The FP calculations of Lee (1992) showed that the rotation parameter
$V/\sigma$ increases as one moves in toward the center of a system with an
empty loss cone (i.e. maximally accreting) black hole.  He found that the
rotation speed increases all the way in to the core, and that the velocity
dispersion is ``Keplerian'' (i.e. $\sigma \sim r^{-1/2}$) over the inner
region.  Our simulations concur with two of these conclusions.  In Figure
\ref{vsig} we show $V$, $\sigma$, and $V/\sigma$ versus radius for four
different times in the simulation XII.  While the dispersion falls as
$r^{-1/2}$, and the rotation speed rises into the center, the behavior of
$V/\sigma$ is not clear.  It appears that the slope of $V/\sigma$ is
increasing as the system evolves, but whether or not it would evolve to the
stage where it rises all the way in to the center is unknown.  We concur
with Lee's result that $V/\sigma$ fails to rise inward for black hole
systems with no loss cone.  However, we found that $V/\sigma$ fails to rise
inward for the mildly rotating model XIVa.  It is possible that this is a
result of limited dynamic range.  While \Mbh/m$_{\star}$ was {\it set} to
$10^6$ for the process of black hole accretion, it is only around 300
{\it gravitationally}, so rotation speeds near the hole are lower than they
would be in a system with N=$10^6$.  If this is indeed the problem, there is
no clear way around it except for increasing N considerably beyond the range
$30 \leq N \leq 3000$ explored in this study.

\section{Conclusions}

Several important pieces of information can be extracted from the simulations
performed in this study.  The first is that a system with the same time scale
ratio $t_{ch}/t_{rh}$ and flattening as M32 has a strong tendency to
evolve away from its initial configuration through stellar merger events and
two-body relaxation.  Specifically, such a system would rapidly build up a
considerably flatter stellar mass spectrum in the core.  Another is that the
dominant merger remnant mass for (physically) collisional systems increases
with increasing N.  Finally, the decision of whether or not to recycle mass
ejecta into stars does not significantly change these trends, although it
does alter the evolution of the core structure.  Taken together, these points
suggest that M32 harbors a black hole in its nucleus.  If the nucleus of M32
contained no central black hole, these simulations show that it would evolve
away from this state rather rapidly.  If the core were composed of stars,
physical stellar collisions would produce a steeper core density profile by
flattening the stellar mass spectrum.  In addition, these collisions would
tend to produce a single dominant merger remnant, further changing the core
structure.  If the core were a cluster of compact objects, energy losses by
gravitational radiation might result in mergers and core collapse (Quinlan
and Shapiro 1989).  Goodman and Lee (1989) put a lower limit of $0.''1$ on the
half-mass radius of any embedded dark cluster, while van der Marel \etal\
(1997) have ruled out dark clusters with a Plummer scale length greater than
$0.''06$.  In either case, our present view of M32 would be of a special point
in its history, at the threshold of an epoch of intense nuclear activity.  The
third scenario, in which the central mass concentration is in the form of a
black hole, would result in a less rapid central evolution, and thus would not
invoke an assumption of our observations being conducted during a special time
period.

An extrapolation of the results of these numerical experiments suggest that
globular clusters are probably incapable of forming black holes through
stellar collisions, whereas a larger system like M32 can form a 100 \Msun
black hole, consistent with Quinlan and Shapiro (1989).  Once formed, all
subsequent collisions serve only to increase the black hole mass, and thus
the black hole formed in this manner can provide the seed required in many
models of active galactic nuclei.

Our simulations concur with the Fokker-Planck of Cohn and Kulsrud (1978)
with regard to the density profile $\rho \propto r^{\alpha}$ in the
vacinity of a central black hole.  For spherical clusters we find that
$\alpha=-1.79 \pm 0.08$, consistent with the Cohn and Kulsrud value of -1.75,
for systems containing equal-mass stars.  Furthermore, we show that the
profile steepens somewhat for rotationally flattened systems, and
significantly for systems following a Salpeter mass spectrum.  In systems
containing a black hole, a Salpeter mass spectrum, stellar collisions and
stellar evolution, we find $\alpha=-2.88\pm 0.16$.

Future work along these lines can be conducted according to two different
guiding principles.  One of these need hardly be stated: to improve the
statistics and the dynamic range, one should increase the number of
particles.  It is essential that the scaling of dominant merger remnant mass
with N be fully understood to extrapolate these results in a more quantitative
fashion to elliptical galaxies.  Performing these experiments with $10^4$ and
$10^5$ particles would produce a more robust scaling relation.  One would also
like to verify the minimum value of N derived in this study for which a system
described by $t_{col}/t_{rel}=100$ will produce a large black hole through
stellar mergers.  The second approach is to combine the results of N$_s$
experiments which differ only in the random variables used to generate the
stellar distribution function (Giersz and Heggie 1994, 1996).  The results
can then be combined in an appropriate manner to extract a higher
signal-to-noise ratio.  With increased N or N$_s$ one can construct
distribution functions in $E$ and $L_z$ which are much more smooth than the
current study would produce.  This would facilitate more detailed and
quantitative comparisons with FP results.

\acknowledgments{This research was supported by NASA Theory Grant NAG 5-2758.
DR thanks the J.\ S.\ Guggenheim Foundation for a fellowship and the Ambrose
Monell Foundation for support.}

\clearpage
\begin{center}
REFERENCES
\end{center}

\def\ARAA{{\it Ann. Rev. Astron. Astrophys.}}
\def\ApJ{{\it Ap. J.}}
\def\ApJL{{\it Ap. J. Lett.}}
\def\ApJSS{{\it Ap. J. Supp. Ser.}}
\def\AandA{{\it Astron. Astrophys.}}
\def\AJ{{\it Astron. J.}}
\def\JCP{{\it J. Comp. Phys.}}
\def\MNRAS{{\it M. N. R. A. S.}}
\def\N{{\it Nature}}
\def\PASJ{{\it Publ. Astron. Soc. Jap.}}
\def\RPP{{\it Rep. Prog. Phys.}}

\begin{verse}

Applegate, J. H. 1986, \ApJ, {\bf 301}, 132.

Barnes, J., and Hut, P. 1986, \N, {\bf 324}, 446.

Benz, W., and Hills, J. G. 1992, \ApJ, {\bf 389}, 546.

Benz, W., and Hills, J. G. 1987, \ApJ, {\bf 323}, 614.

Bethe, H. A. 1986, in {\it Highlights of Modern Astrophysics}, ed. S. L.
   Shapiro and S. A. Teukolsky (New York: Wiley), p. 45.

Binney, J. J., Davies, R. L., and Illingworth, G. D. 1990, \ApJ {\bf 361}, 78.

Binney, J. J., and Tremaine, S. 1987, {\it Galactic Dynamics} (Princeton:
Princeton University Press).

Bodenheimer, P., and Woosley, S. E. 1983, \ApJ, {\bf 269}, 281.

Bond, J. R., Arnett, W. D., and Carr, B. J. 1984, \ApJ, {\bf 280}, 825.

Chernoff, D. F., and Weinberg, M. D. 1990, \ApJ, {\bf 351}, 121.

Chitre, D. M., and Hartle, J.B. 1976,  \ApJ, {\bf 207}, 592.

Cohn, H. N., and Kulsrud, R. M. 1978, \ApJ {\bf 226}, 1087.

Cowley, A. P. 1992, \ARAA, {\bf 30}, 287.

David, L. P., Durisen, R. H., and Cohn, H. N. 1987, \ApJ, {\bf 316}, 505.

Dejonghe, H., and De Zeeuw, T. 1988, \ApJ, {\bf 333}, 90.

Freedman, W. L. 1989, \AJ, {\bf 98}, 1285.

Giersz, M., and Heggie, C. D. 1994, \MNRAS, {\bf 268}, 257.

Giersz, M., and Heggie, C. D. 1996, \MNRAS, {\bf 279}, 1037.

Goodman, J. and Lee, H. M. 1989 \ApJ, {\bf 337}, 84.

Grossman, A. S., Hays, D., and Graboske, H. C. 1974, \AandA, {\bf 30}, 95.

H\'{e}non, M. 1975, in {\it IAU Symp. No. 69, Dynamics of Stellar Systems},
ed. A. Hayli (D. Reidel Publishing Company, Dordrecht), p. 133.

Hernquist, L. 1987, \ApJSS, {\bf 64}, 715.
 
Hernquist, L. 1990, \JCP, {\bf 87}, 137.

Hulse, R. A., Taylor, J. H. 1975, \ApJL, {\bf 195}, L51.

Humphreys, R. M., and Davidson, K. 1986, {\it New Scientist}, {\bf 112}, 38.

Iben, I. 1964, \ApJ, {\bf 140}, 1631.

Iben, I. 1967, \ApJ, {\bf 147}, 624.

Iben, I., and Renzini, A. 1983, \ARAA, {\bf 21}, 271.

Kalnajs, A. J. 1972, \ApJ, {\bf 175}, 63.

Kippenhahn, R., and Weigert, A. 1990, {\it Stellar Structure and Evolution}
(New York: Springer-Verlag).

Kuzmin, G. G., and Kutuzov, S. A. 1962, {\it Bull. Abastumani Ap. Obs.},
{\bf 27}, 82.

Lai, D., Rasio, F. A., and Shapiro, S. L. 1993, \ApJ, {\bf 412}, 593.

Larson, R. B. 1977, in {\it The Evolution of Galaxies and Stellar Populations},
ed. B. M. Tinsley and R. B. Larson (New Haven: Yale University Observatory),
p. 97.

Lauer, T., Faber, S. M., Lynds, C. R., Baum, W. A., Ewald, S. P., Groth,
E. J., Hester, J. J., Holtzman, J. A., Kristian, J., and Light, R. M. 1992,
\AJ, {\bf 103}, 703.

Lee, M. H. 1992, Ph.D. Thesis, Princeton University.

Mathews, W. G. 1972, \ApJ, {\bf 174}, 101.

Ostriker, J. P., and Peebles, P. J. E. 1973, \ApJ, {\bf 186}, 467.

Quinlan, G. D., Hernquist, L., and Sigurdsson, S. 1995, \ApJ, {\bf 440}, 554.

Quinlan, G. D., and Shapiro, S. L. 1990, \ApJ, {\bf 356}, 483.

Quinlan, G. D., and Shapiro, S. L. 1989, \ApJ, {\bf 343}, 725.

Reimers, D., and Koester, D. 1982, \AandA, {\bf 116}, 341.

Richstone, D. O., Bower, G., and Dressler, A. 1990, \ApJ, {\bf 353}, 118.

Salpeter, E. E. 1955, \ApJ, {\bf 121}, 161.

Sanders, R. H. 1970, \ApJ, {\bf 162}, 791.

Satoh, C. 1980, \PASJ, {\bf 32}, 41.

Spitzer, L. Jr., and Saslaw, W. C., 1966, \ApJ, {\bf 143}, 400.

Strom, R. G. 1988, \MNRAS, {\bf 230}, 331.

Taylor, J. H., and Dewey, R. J. 1988, \ApJ, {\bf 332}, 770.

Taylor, J. H., and Weisberg, J. M. 1989, \ApJ, {\bf 345}, 434.

Tonry, J. L. 1989, in {\it Dynamics of Dense Stellar Systems}, ed. D. Merritt
(Cambridge: Cambridge University Press), p. 35.

van der Marel, R. P., de Zeeuw, P. T., Rix, H. W., and Quinlan, G. D. 1997,
\N, {\bf 385}, 610.

Weidemann, V. 1990, \ARAA, {\bf 28}, 103.

Woosley, S. E., and Weaver, T. A. 1986, \ARAA, {\bf 24}, 205.

Young, P. 1980, \ApJ {\bf 242}, 1232.

\end{verse}


\clearpage

\begin{deluxetable}{lcc}
\tablecaption{Four Galactic and Magellanic black hole candidates.
\label{bhcands}}
\tablehead{
\colhead{candidate} &
\colhead{lower mass limit (\Msun)} &
\colhead{probable mass (\Msun)}
}
\startdata
   Cygnus X-1   &    6               & $16\pm 5$      \nl
   LMC X-3      &    3               & $> 9$          \nl
   LMC X-1      &    4               & $7\pm 3$       \nl
   0620-00      &    3               & $> 7.3$        \nl
\enddata
\end{deluxetable}

\clearpage

\begin{deluxetable}{ccrccccccc}
\tablecaption{
Initial model parameters for the simulations in this study. \label{modpars}}
\tablehead{
\colhead{Model} &
\colhead{$c/a$} &
\colhead{N} &
\colhead{IMF/\Msun} &
\colhead{range/\Msun} &
\colhead{$\star$ evo.} &
\colhead{$\star$ col.} &
\colhead{$\star$ for.} &
\colhead{$\frac{t_{col}}{t_{rel}}$} &
\colhead{\Mbh/M}
}
\startdata
Group 1 &   &      &       &       &   &   &   &     &     \nl
XVIa  & 1.0 &  300 & \imfa & \rana & n & y & n & 100 &  0  \nl
XVIb  & 0.9 &  300 & \imfa & \rana & n & y & n & 100 &  0  \nl
XVIc  & 0.8 &  300 & \imfa & \rana & n & y & n & 100 &  0  \nl
XVId  & 0.7 &  300 & \imfa & \rana & n & y & n & 100 &  0  \nl
XVIe  & 0.6 &  300 & \imfa & \rana & n & y & n & 100 &  0  \nl
XVIf  & 0.5 &  300 & \imfa & \rana & n & y & n & 100 &  0  \nl
XVIg  & 0.4 &  300 & \imfa & \rana & n & y & n & 100 &  0  \nl
XVIh  & 0.3 &  300 & \imfa & \rana & n & y & n & 100 &  0  \nl
Group 2 &   &      &       &       &   &   &   &     &     \nl
  IX  & 1.0 & 3000 & \imfa & \rana & y & y & n & 100 &  0  \nl
   X  & 0.3 & 3000 & \imfa & \rana & y & y & n & 100 &  0  \nl
  XI  & 1.0 & 3000 & \imfa & \rana & y & y & n & 100 & 0.1 \nl
 XII  & 0.3 & 3000 & \imfa & \rana & y & y & n & 100 & 0.1 \nl
Group 3 &   &      &       &       &   &   &   &     &     \nl
XIIIa & 0.8 & 3000 & \imfb & \ranb & y & y & n & 100 &  0  \nl
XIIIb & 0.8 & 1000 & \imfb & \ranb & y & y & n & 100 &  0  \nl
XIIIc & 0.8 &  300 & \imfb & \ranb & y & y & n & 100 &  0  \nl
XIIId & 0.8 &  100 & \imfb & \ranb & y & y & n & 100 &  0  \nl
XIIIe & 0.8 &   30 & \imfb & \ranb & y & y & n & 100 &  0  \nl
 XIVa & 0.8 & 3000 & \imfb & \ranb & y & y & n & 100 & 0.1 \nl
 XIVb & 0.8 & 1000 & \imfb & \ranb & n & n & n & 100 & 0.1 \nl
 XV   & 0.8 & 3000 & \imfb & \ranb & y & y & y & 100 &  0  \nl
XVII  & 0.8 & 3000 & \imfb & \ranb & n & n & n & 100 &  0  \nl
XVIII & 0.8 & 3000 & \imfb & \ranb & y & y & y &  10 &  0  \nl
\enddata
\end{deluxetable}

\clearpage

\begin{deluxetable}{crccccl}
\tablecaption{Dominant merger remnants produced in a selection of the
simulations. \label{domstarmass}}
\tablehead{
\colhead{model} &
\colhead{N/2} &
\colhead{$\star$ evo.} &
\colhead{$\star$ col.} &
\colhead{$\star$ for.} &
\colhead{m$_{rem}$/\Msun} &
\colhead{$\epsilon$}
}
\startdata
IX    & 1500 &    yes & yes &   no    &     3.0   &   0.00200  \nl
X     & 1500 &    yes & yes &   no    &     5.0   &   0.00333  \nl
XI    & 1500 &    yes & yes &   no    &     8.0   &   0.00533  \nl
XII   & 1500 &    yes & yes &   no    &    10.0   &   0.00667  \nl
XV    & 1500 &    yes & yes &  yes    &     4.1   &   0.00271  \nl
XVIg  &  150 &     no & yes &   no    &    43.0   &   0.287    \nl
      &      &        &     &         &           &            \nl
XIIIa & 1500 &    yes & yes &   no    &     2.7   &   0.00160  \nl
XIIIb &  500 &    yes & yes &   no    &     2.8   &   0.00560  \nl
XIIIc &  150 &    yes & yes &   no    &     2.4   &   0.0160   \nl
XIIId &   50 &    yes & yes &   no    &     2.2   &   0.0440   \nl
XIIIe &   15 &    yes & yes &   no    &     1.9   &   0.0380   \nl
\enddata
\end{deluxetable}


\clearpage
\plotone{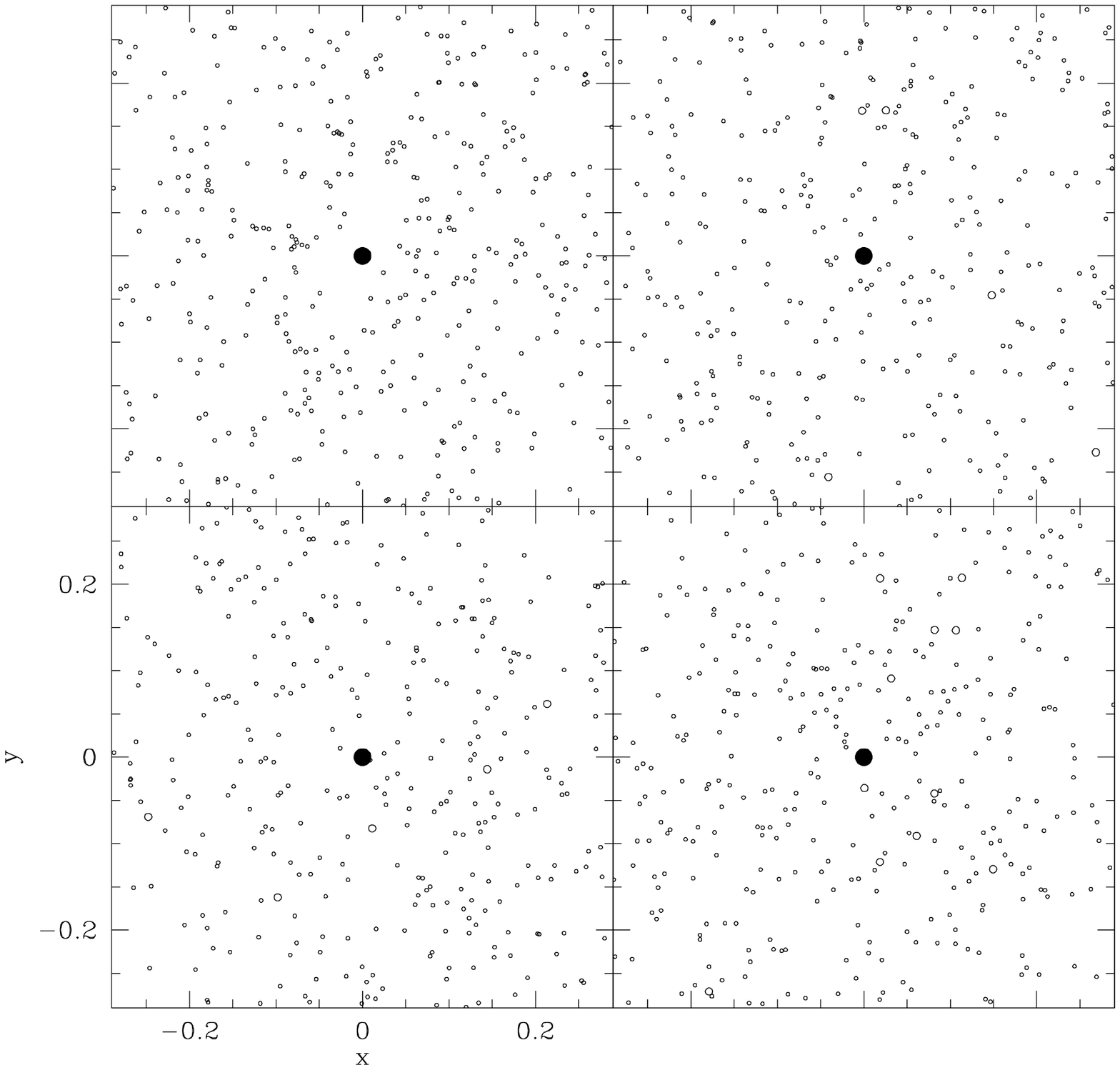}

\figcaption{The early evolution of the central region of a spherical system
containing a black hole (model XI; see Table \ref{modpars}) at $t/t_{rc}=$0,
1, 2, and 3.
\label{cent_rel}}

\clearpage
\plotone{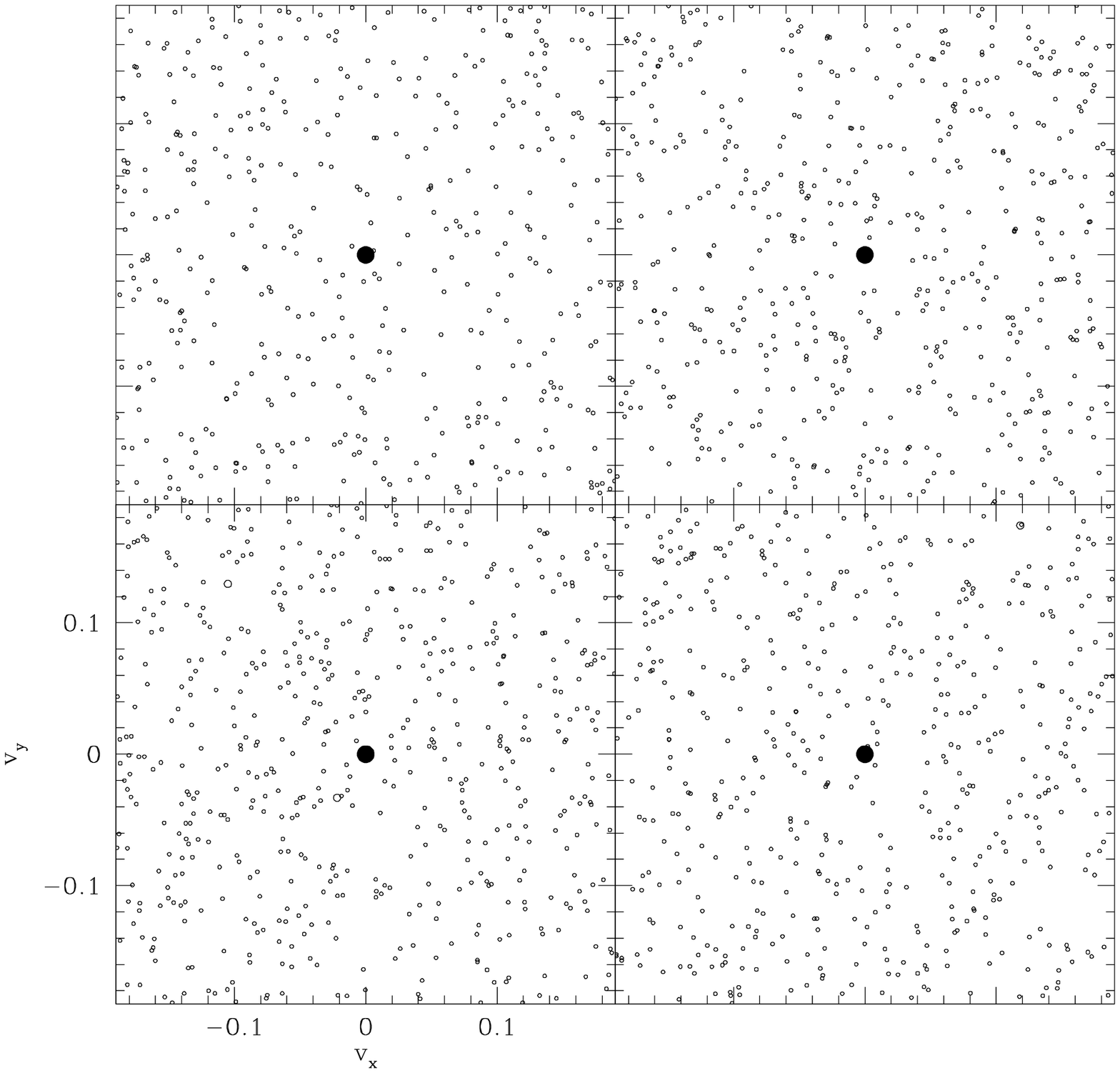}

\figcaption{The early velocity evolution of the center of a spherical system
containing a black hole (model XI; see Table \ref{modpars}) at $t/t_{rc}=$0,
1, 2, and 3.
\label{cent_rel_V}}

\clearpage
\plotone{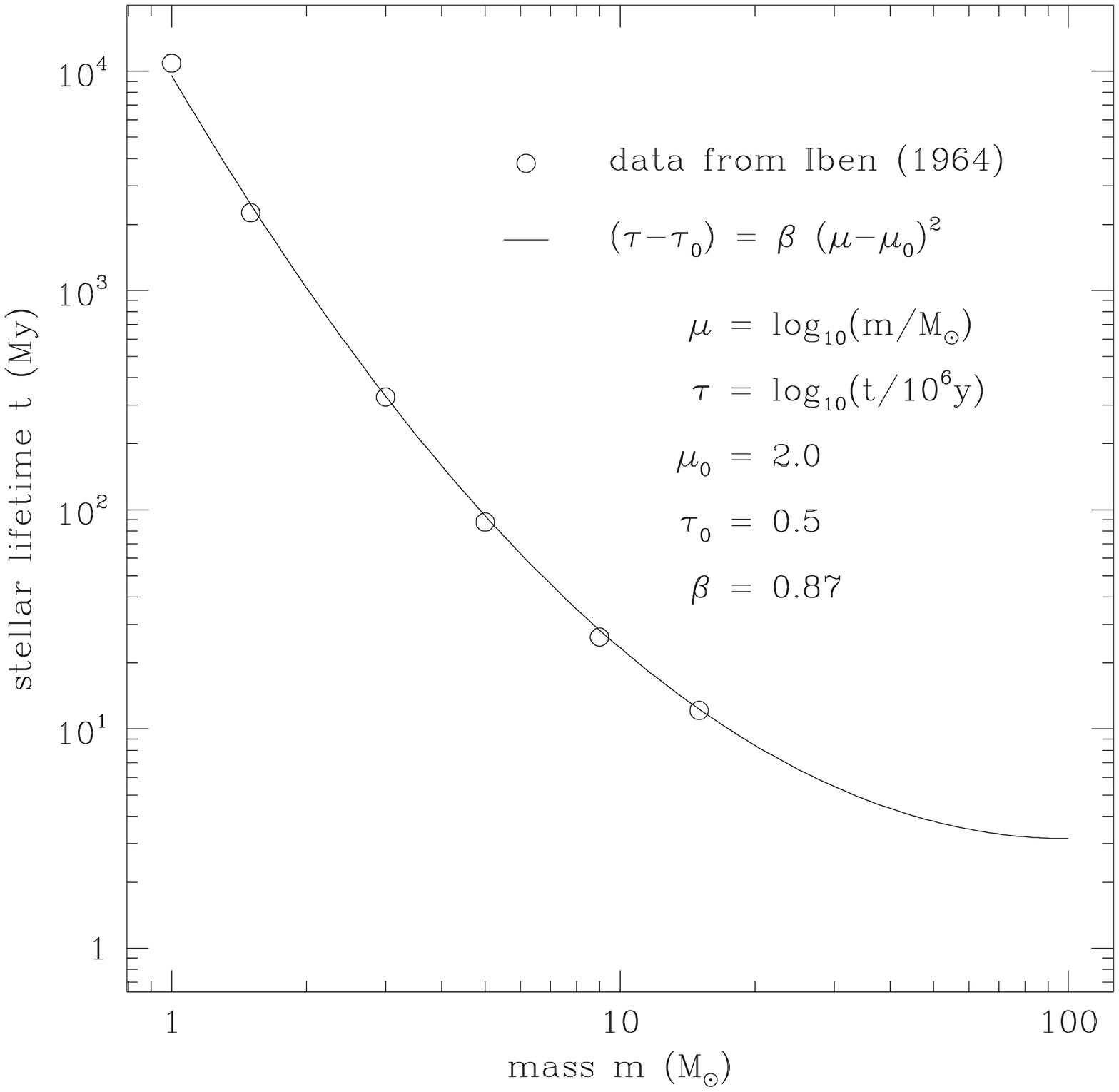}

\figcaption{Main sequence life time versus mass for stars in simulations
incorporating stellar evolution effects.
\label{lifetime}}

\clearpage
\plotone{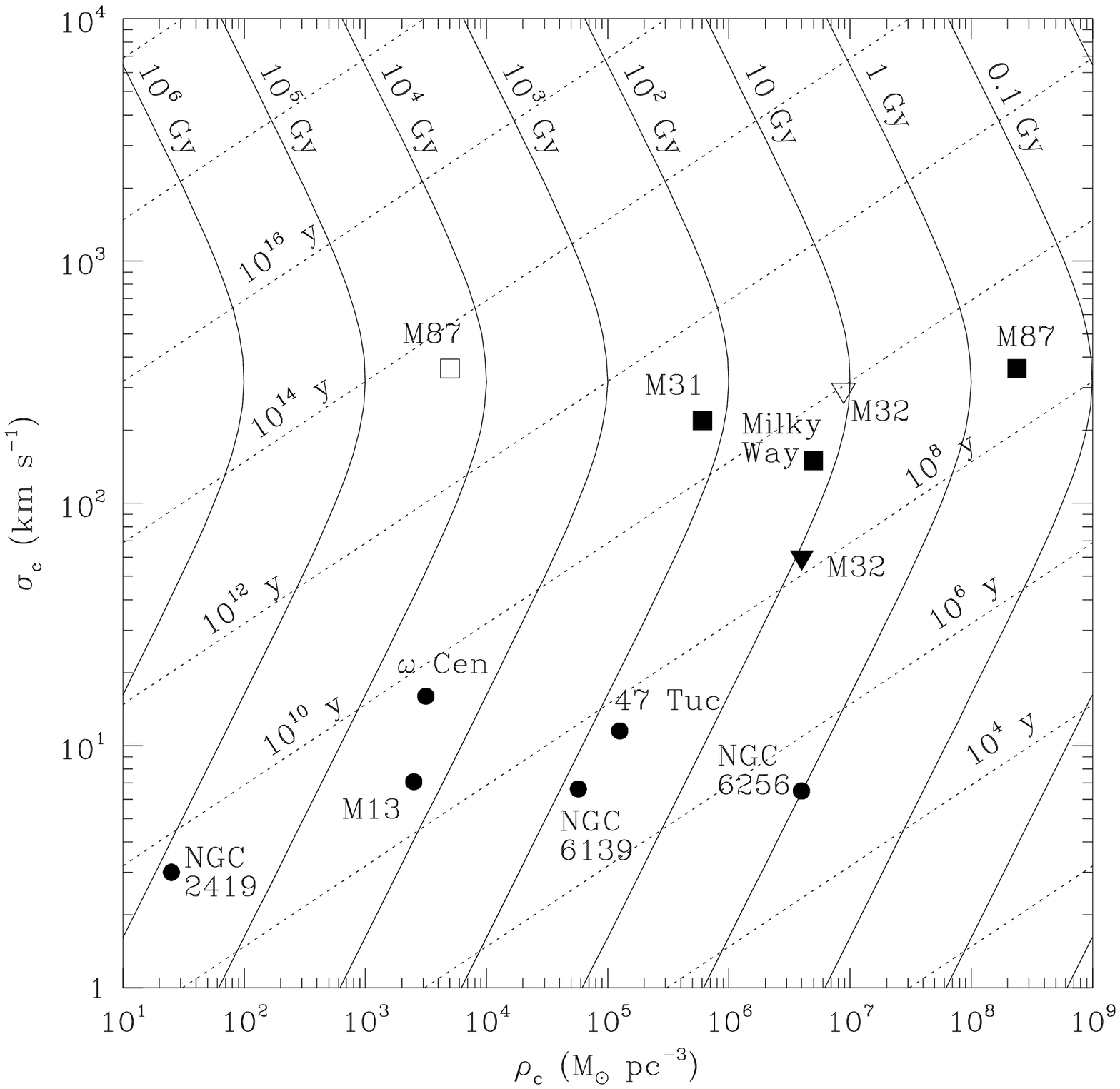}

\figcaption{Stellar collision (solid) and relaxation (dashed) time scale
contours for a large range of core dispersion and density.  The cores of
several local stellar systems are included.  \label{tcol_local}}

\clearpage
\plotone{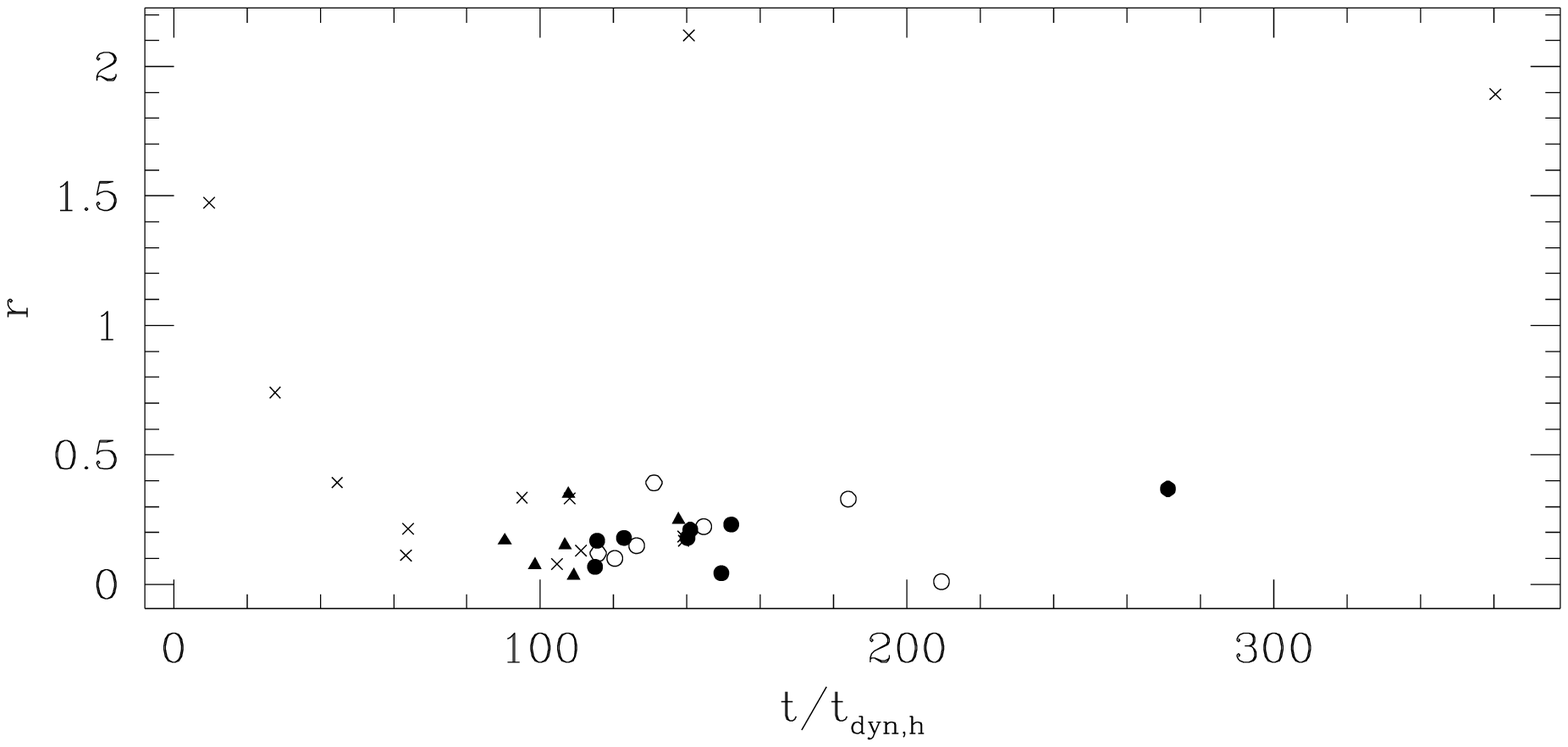}

\figcaption{Center of mass distance of collision events for spherical model
XVIa. \label{treecol}}

\clearpage
\plotone{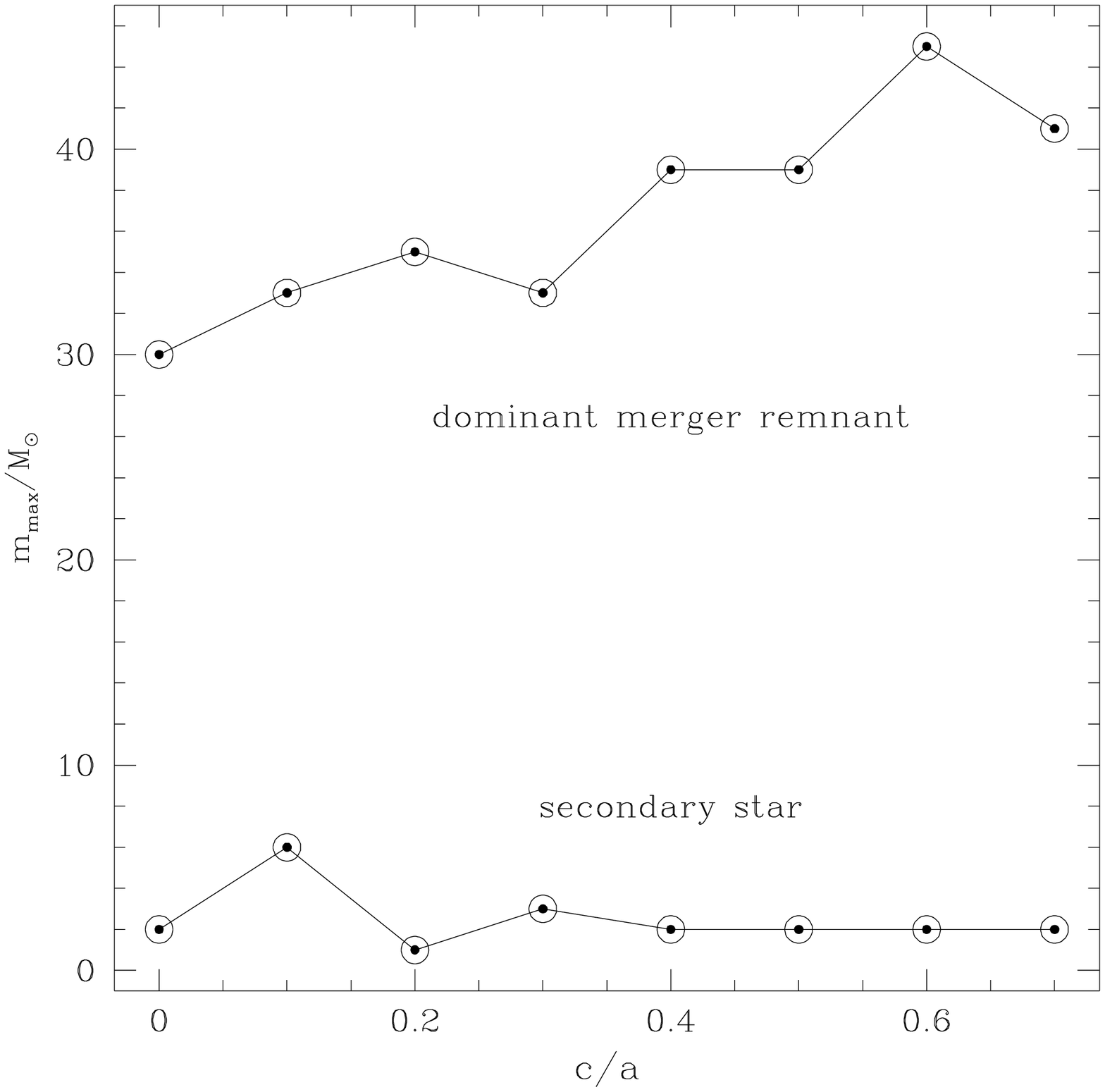}

\figcaption{Dominant merger remnant mass and secondary star mass as a function
of the initial KK model axial ratio.  \label{mmaxeps_cor}}

\clearpage
\plotone{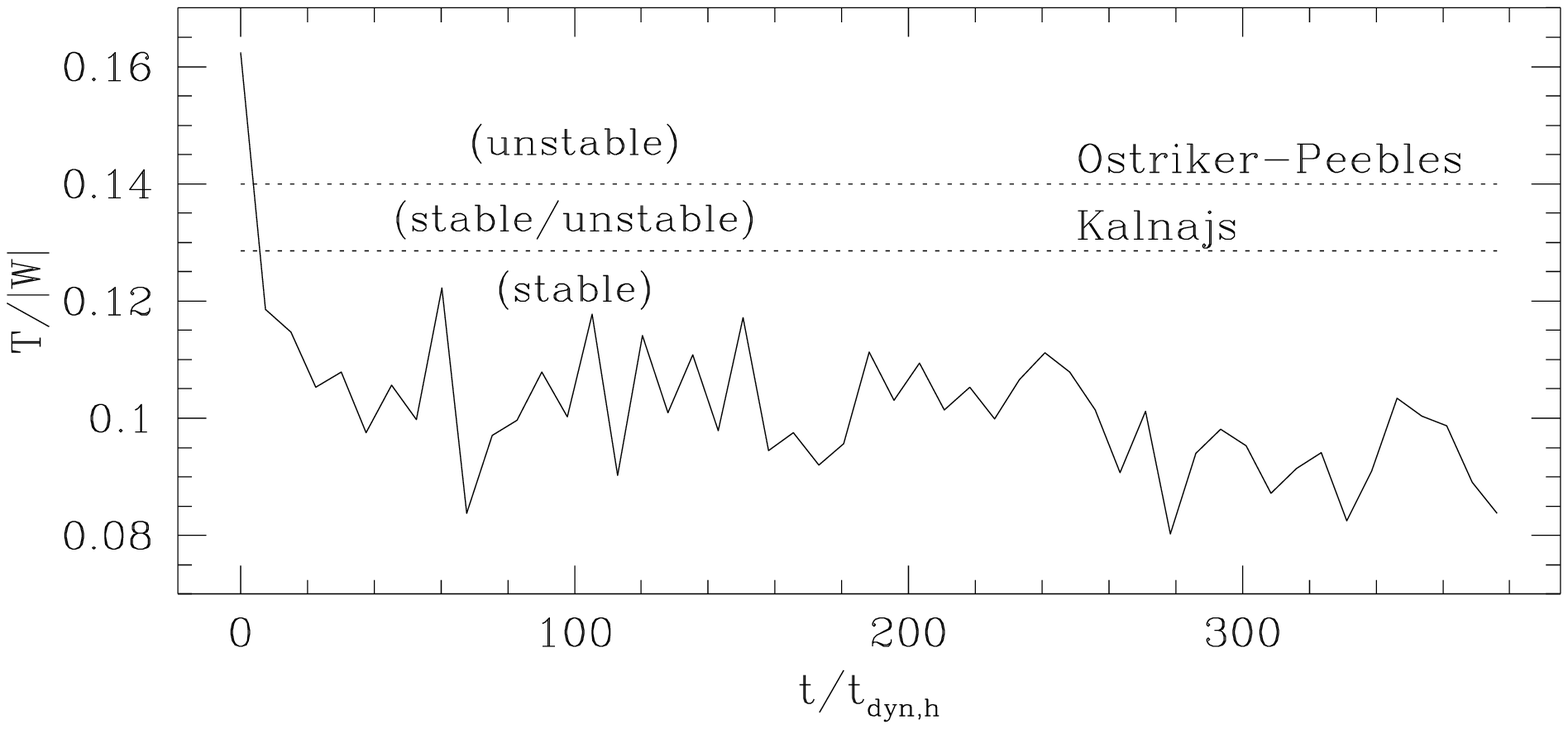}

\figcaption{The $T/|W|$ evolution of model XVIh ($c/a=0.3$, 300 equal-mass
stars). \label{TW}}

\clearpage
\plotone{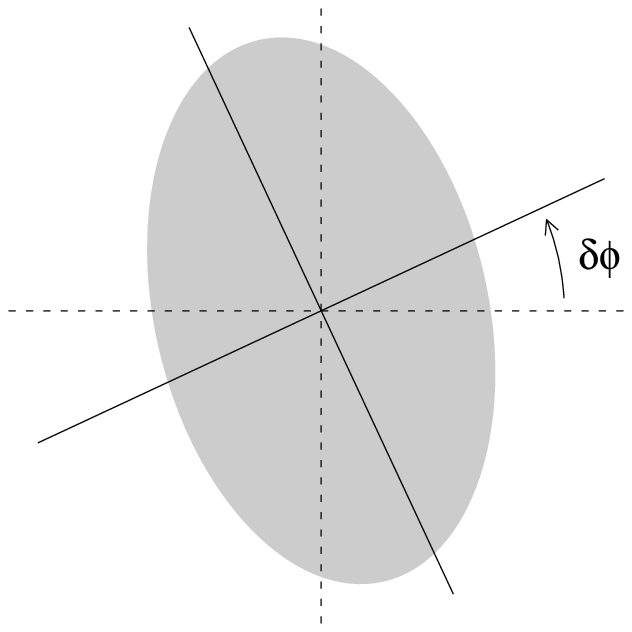}

\figcaption{Geometry for identifying the bar structure of a low-N system.
\label{bar3schem}}

\clearpage
\plotone{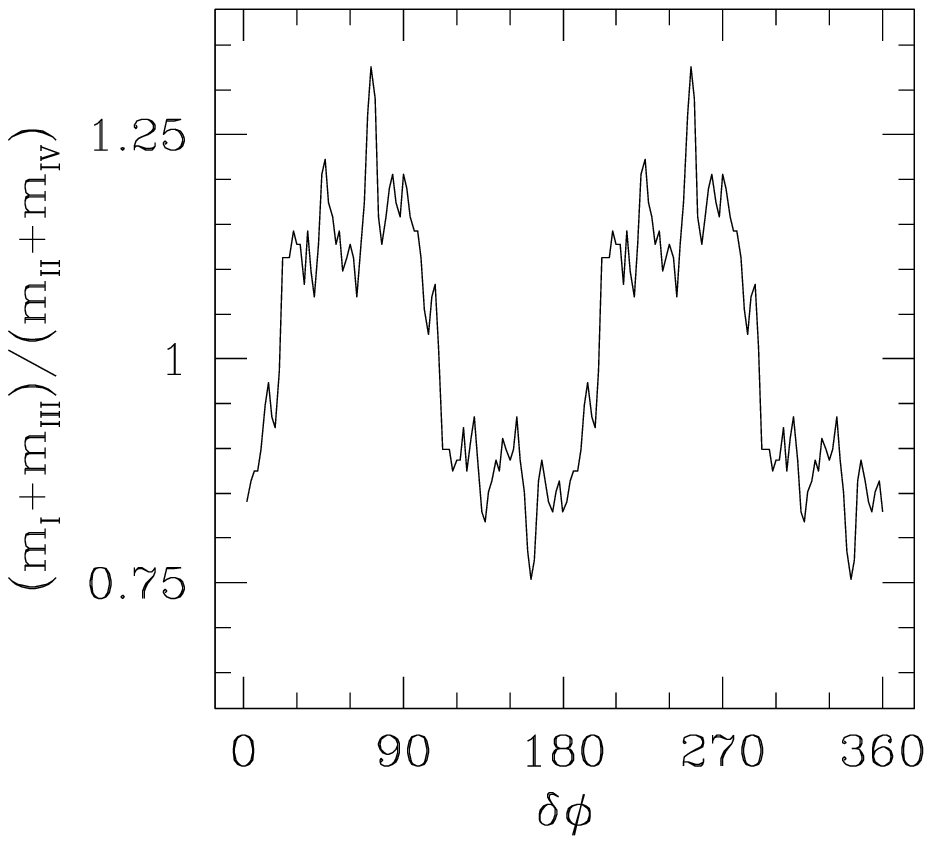}

\figcaption{The ratio of opposite-quadrant mass sums as a function of axial
offset for model XVIh.
\label{bar3}}

\clearpage
\plotone{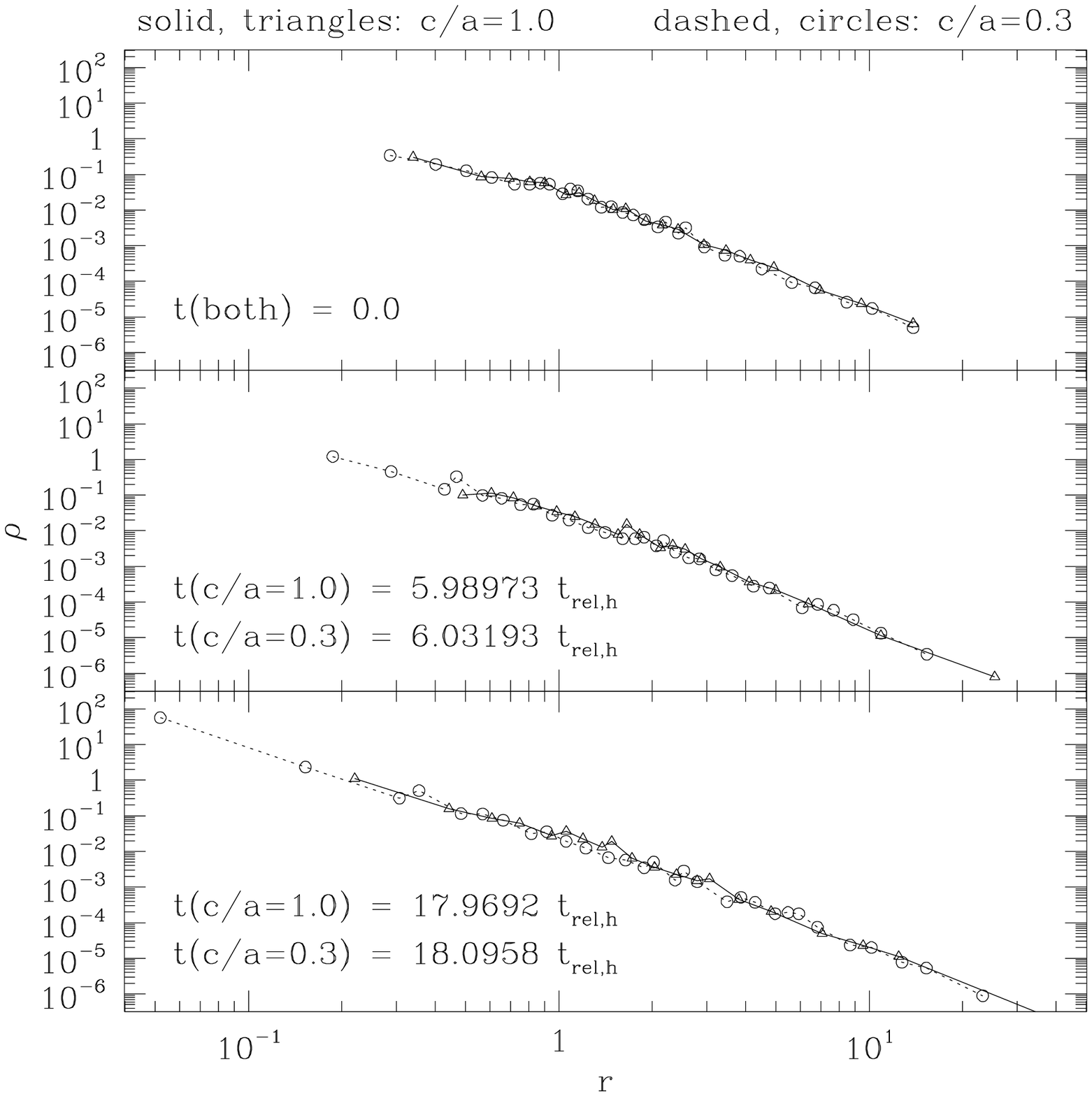}

\figcaption{Density profiles for the spherical and the flattest Group 1
systems.  Note that the times given are in units of the initial half-mass
relaxation time scale.  \label{deps}}

\clearpage
\plotone{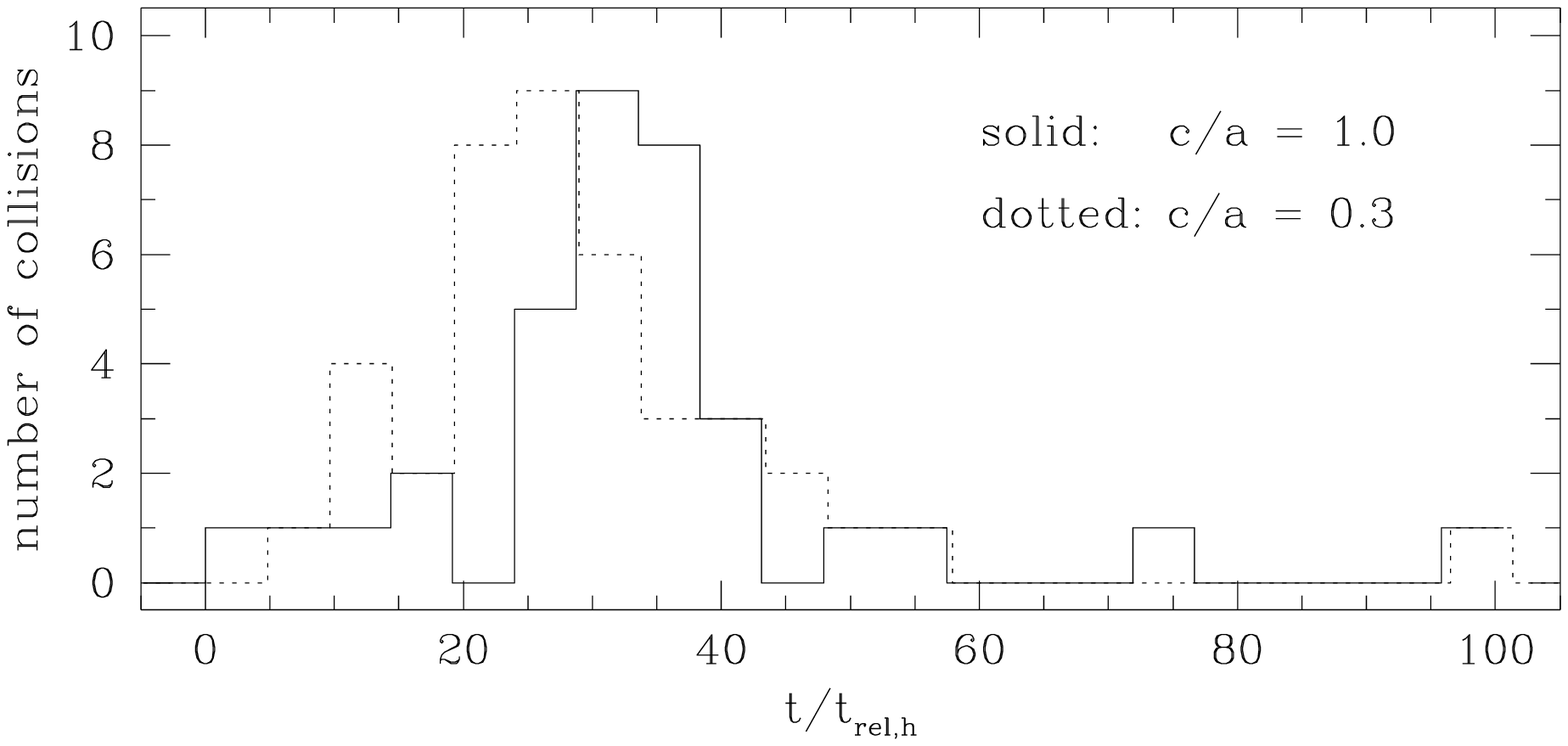}

\figcaption{Collision rate for the spherical and the flattest Group 1
simulations. \label{coll_ip}}

\clearpage
\plotone{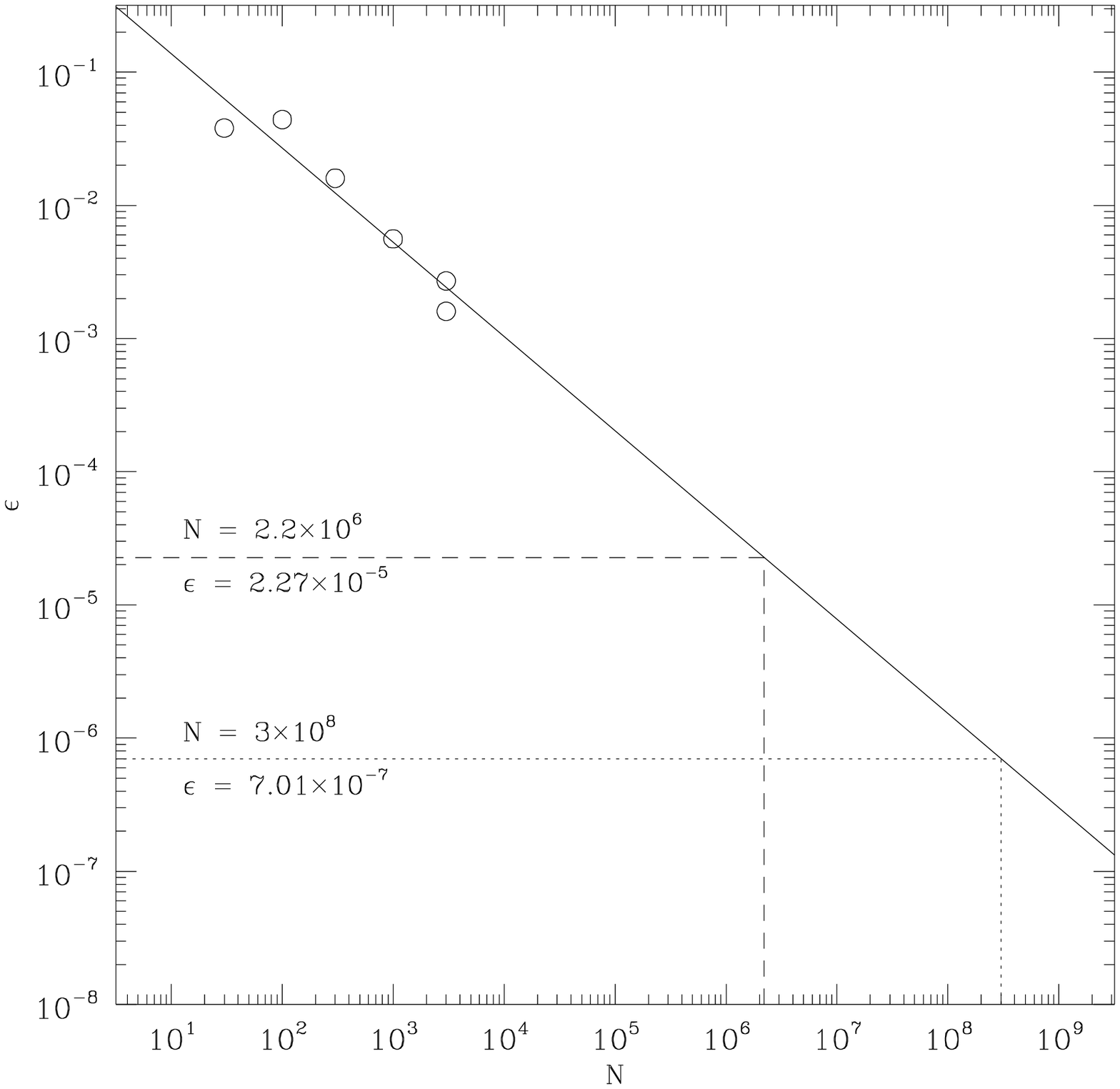}

\figcaption{The stellar merger efficiencies of models XIII and XV.  The
power law fit to the points has a slope of -0.708 and an intercept of -0.155.
\label{mergeeff}}

\clearpage
\plotone{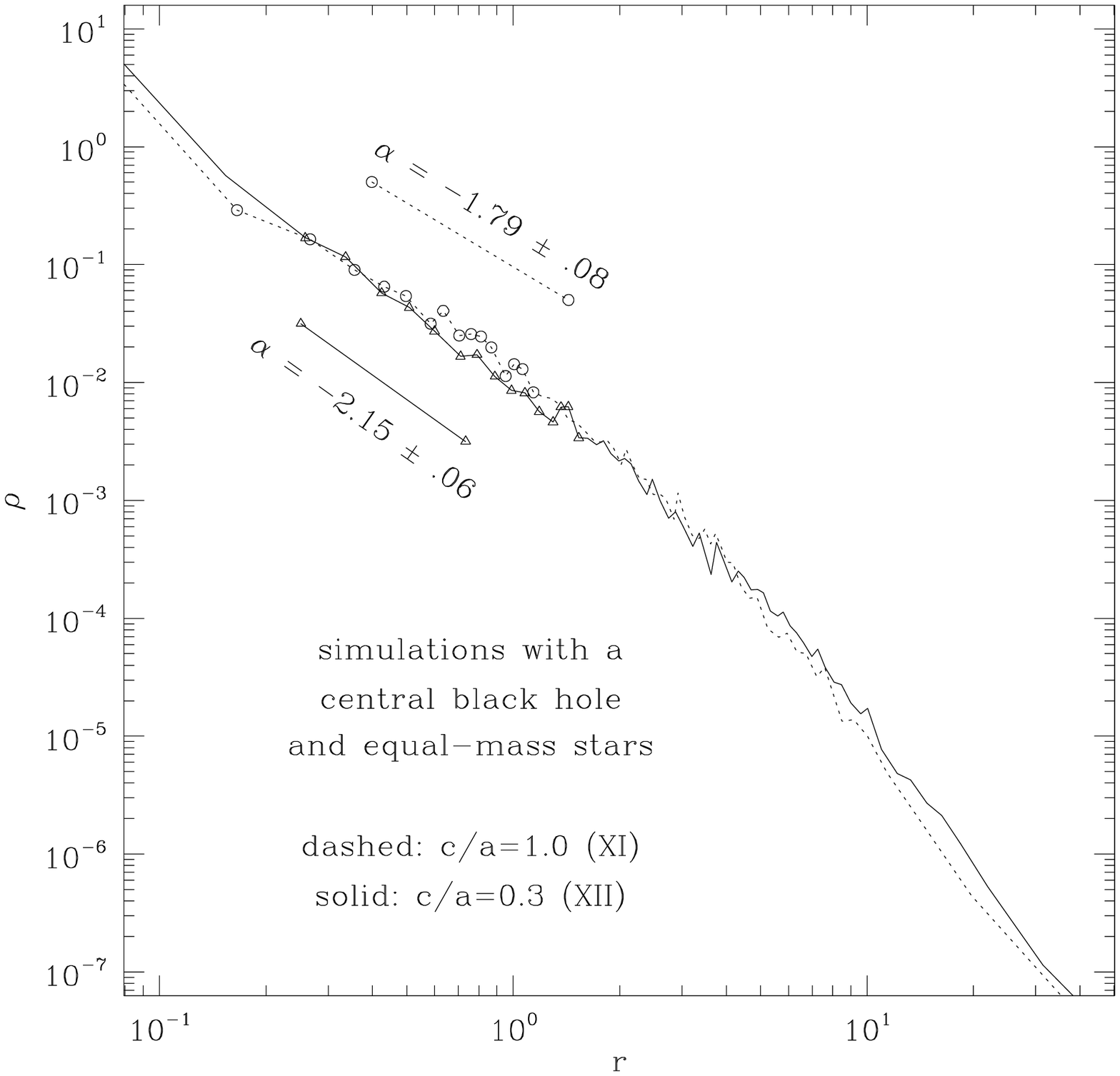}

\figcaption{Density profiles for simulations containg a central black hole
and equal-mass stars.  The dashed line is a spherical model (XI) and the solid
line is rotationally flattened (XII). \label{den_deBH}}

\clearpage
\plotone{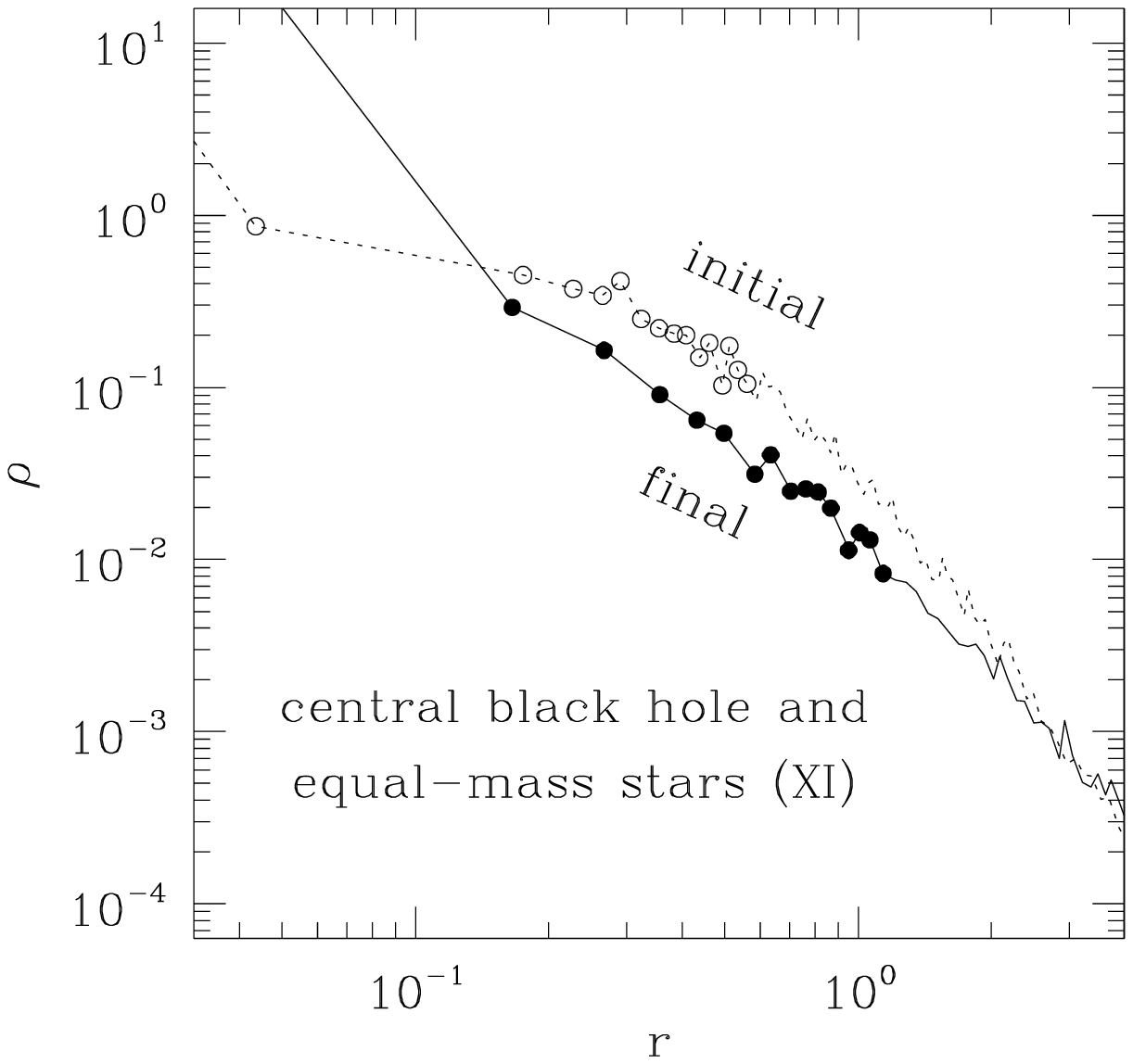}

\figcaption{Initial and final density profiles of model XI.
\label{den_d_evo}}

\clearpage
\plotone{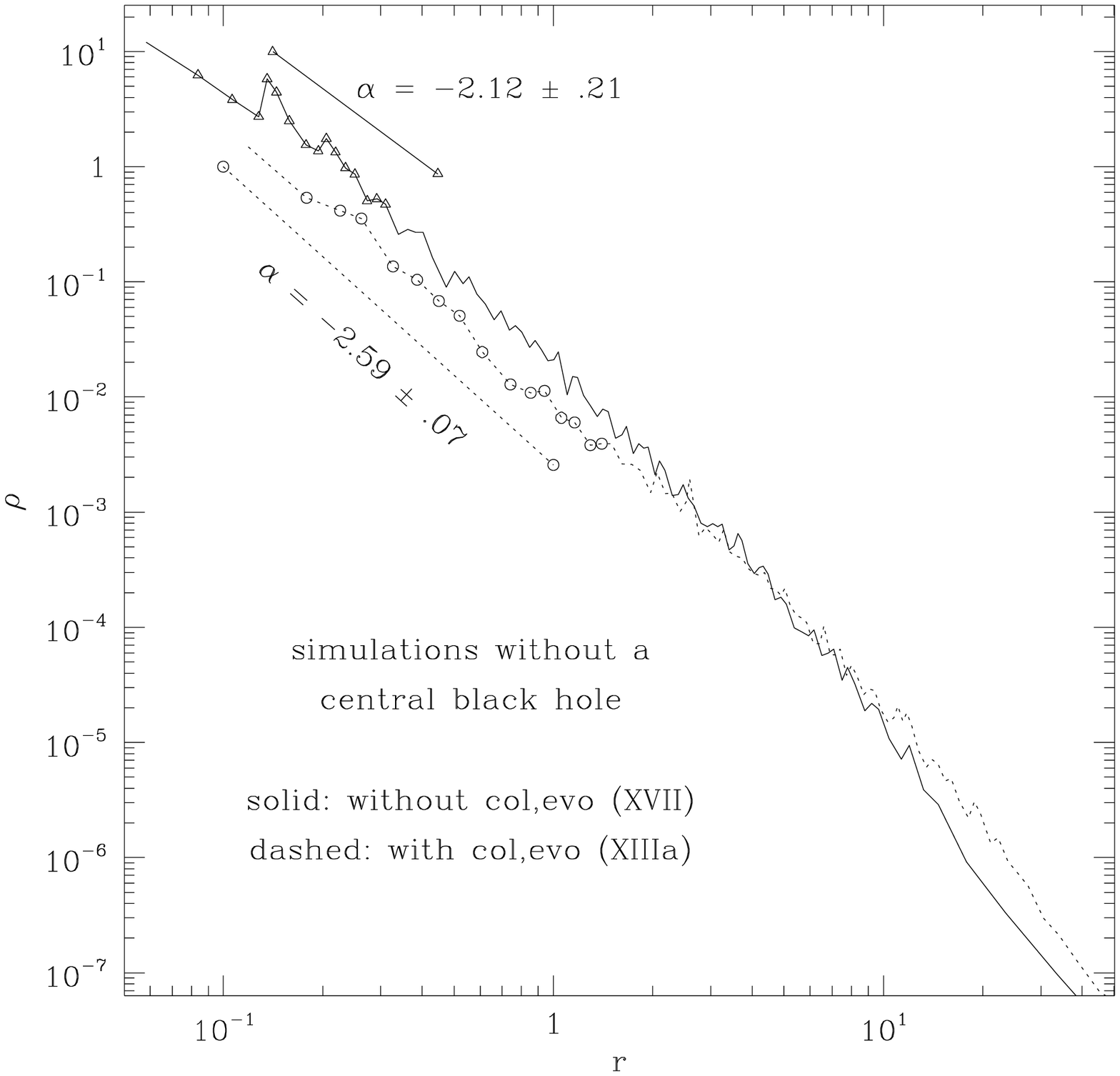}

\figcaption{Density profiles for Salpter IMF $c/a=0.8$ simulations without
central black holes. \label{den_qf}}

\clearpage
\plotone{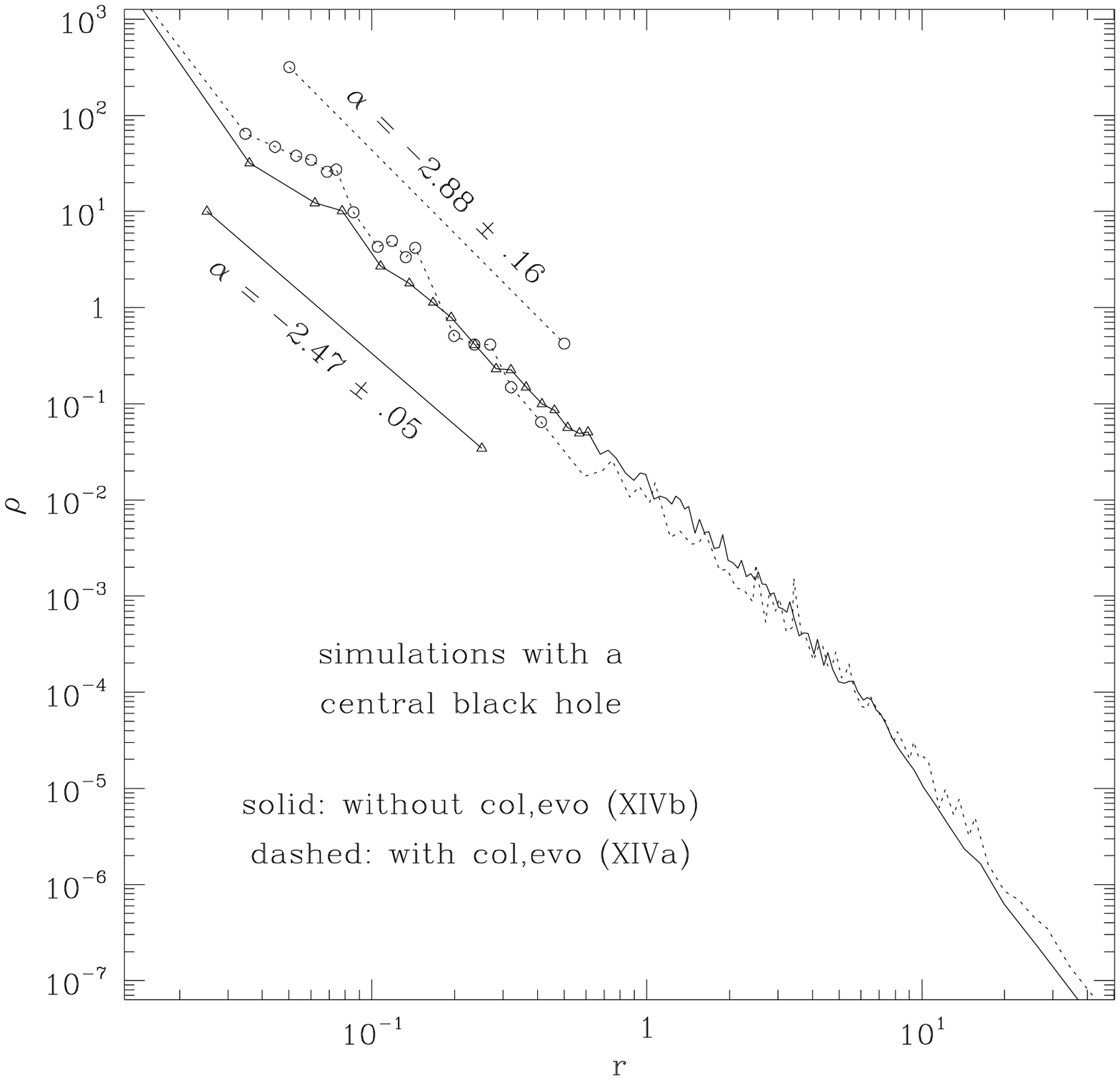}

\figcaption{Density profiles for Salpeter IMF $c/a=0.8$ simulations with
central black holes. \label{den_ggBH}}

\clearpage
\plotone{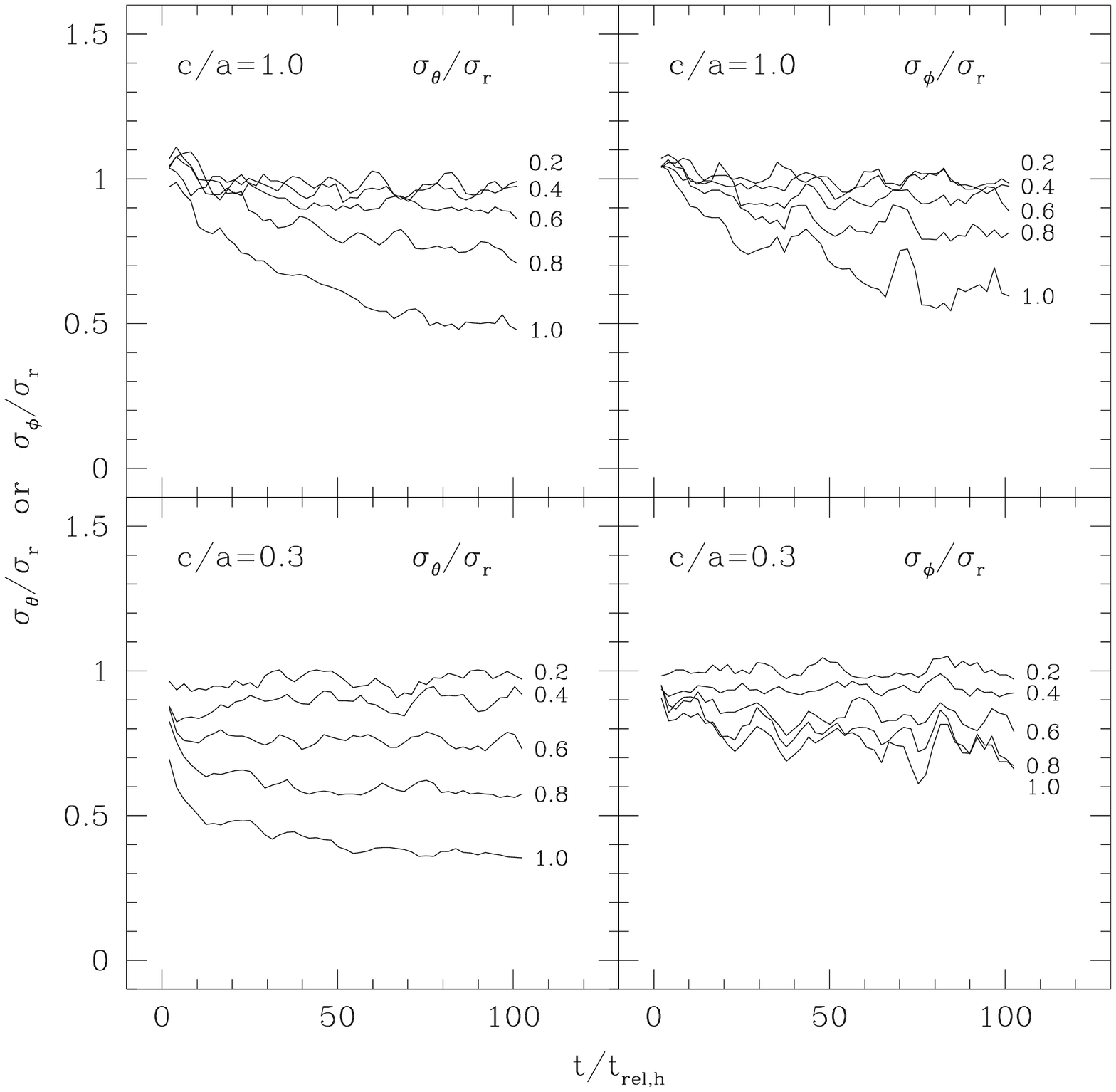}

\figcaption{Developing radial anisotropy in models which contain no central
black hole, for two different rotational states (IX and X). \label{sigrt_ac}}

\clearpage
\plotone{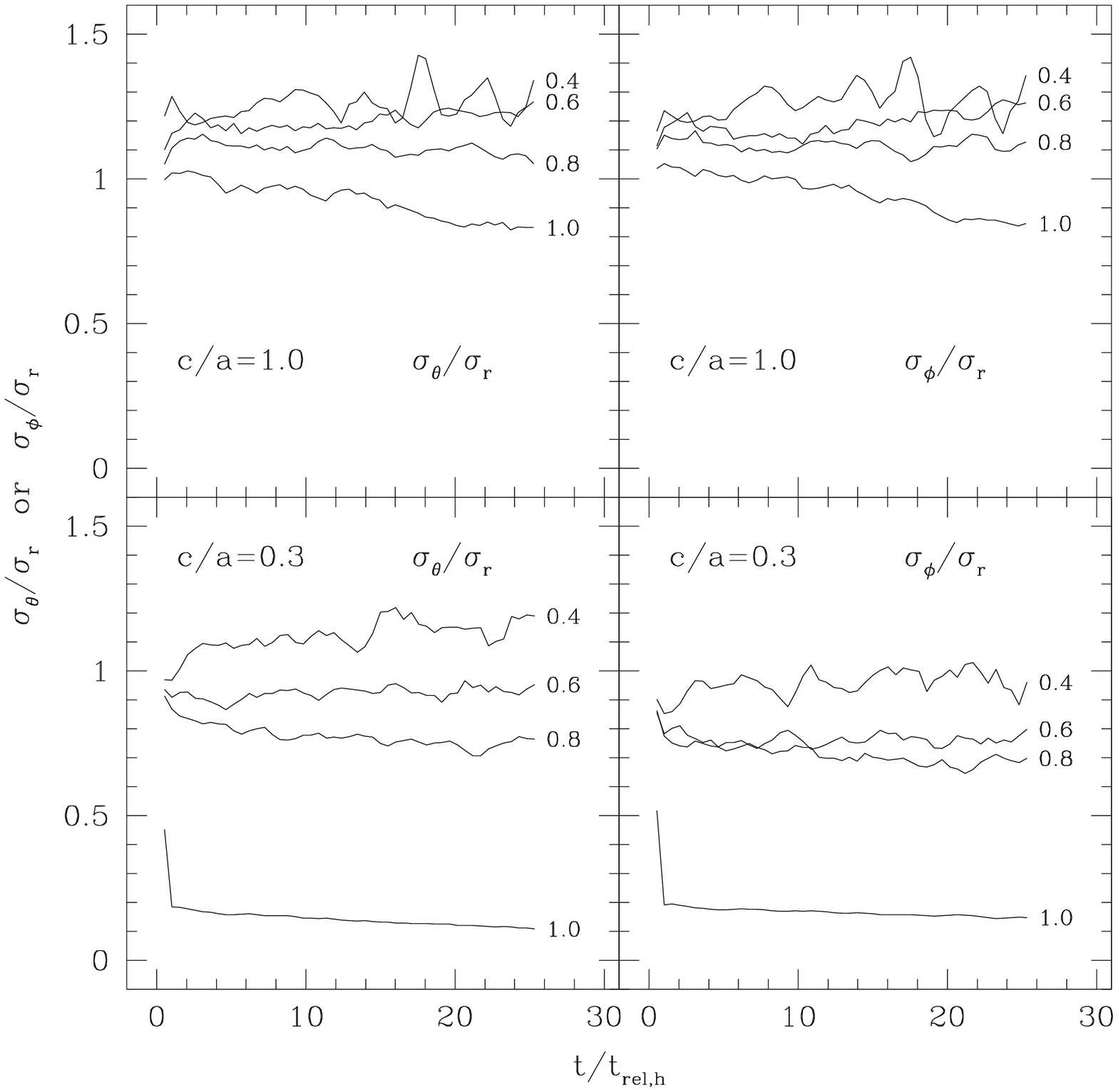}

\figcaption{Developing anisotropy in models XI and XII (equal-mass stars
with a central black hole). \label{sigrt_de}}

\clearpage
\plotone{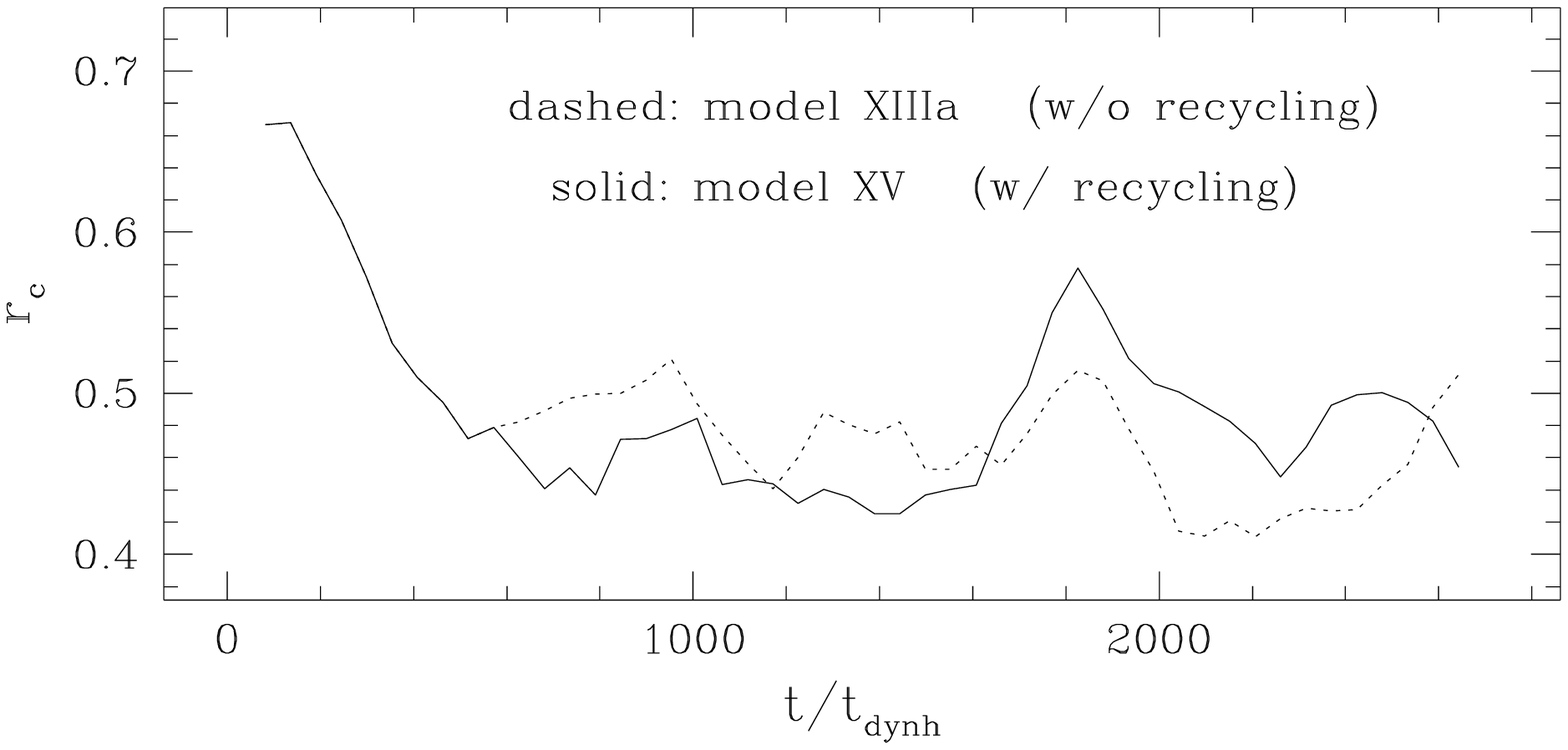}

\figcaption{Comparison of core radius evolution between systems which do
(XV) and do not (XIIIa) recycle stellar ejecta into new stars.
\label{core_r_recy}}

\clearpage
\plotone{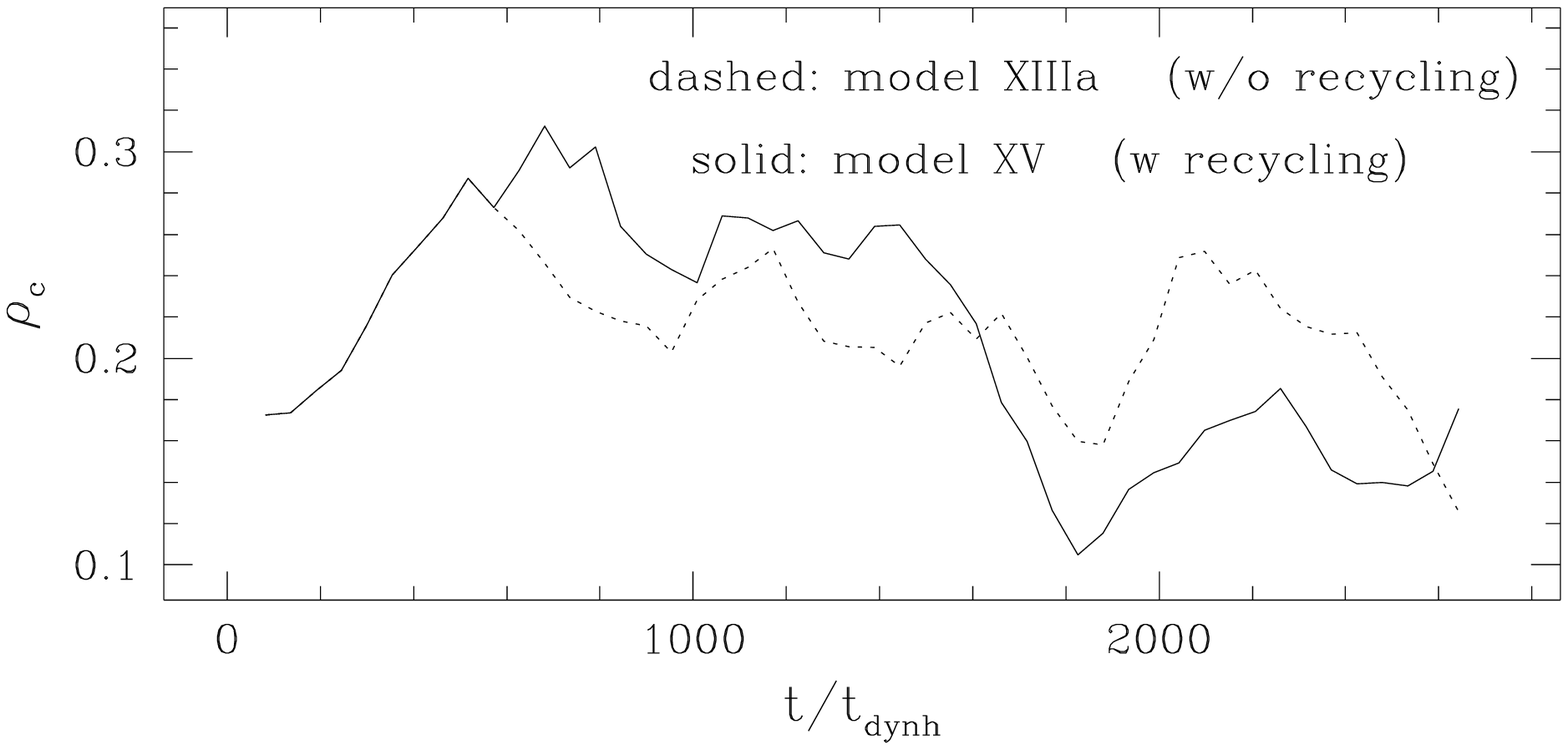}

\figcaption{Comparison of core density evolution between systems which do
and do not recycle stellar ejecta into new stars. \label{core_p_recy}}

\clearpage
\plotone{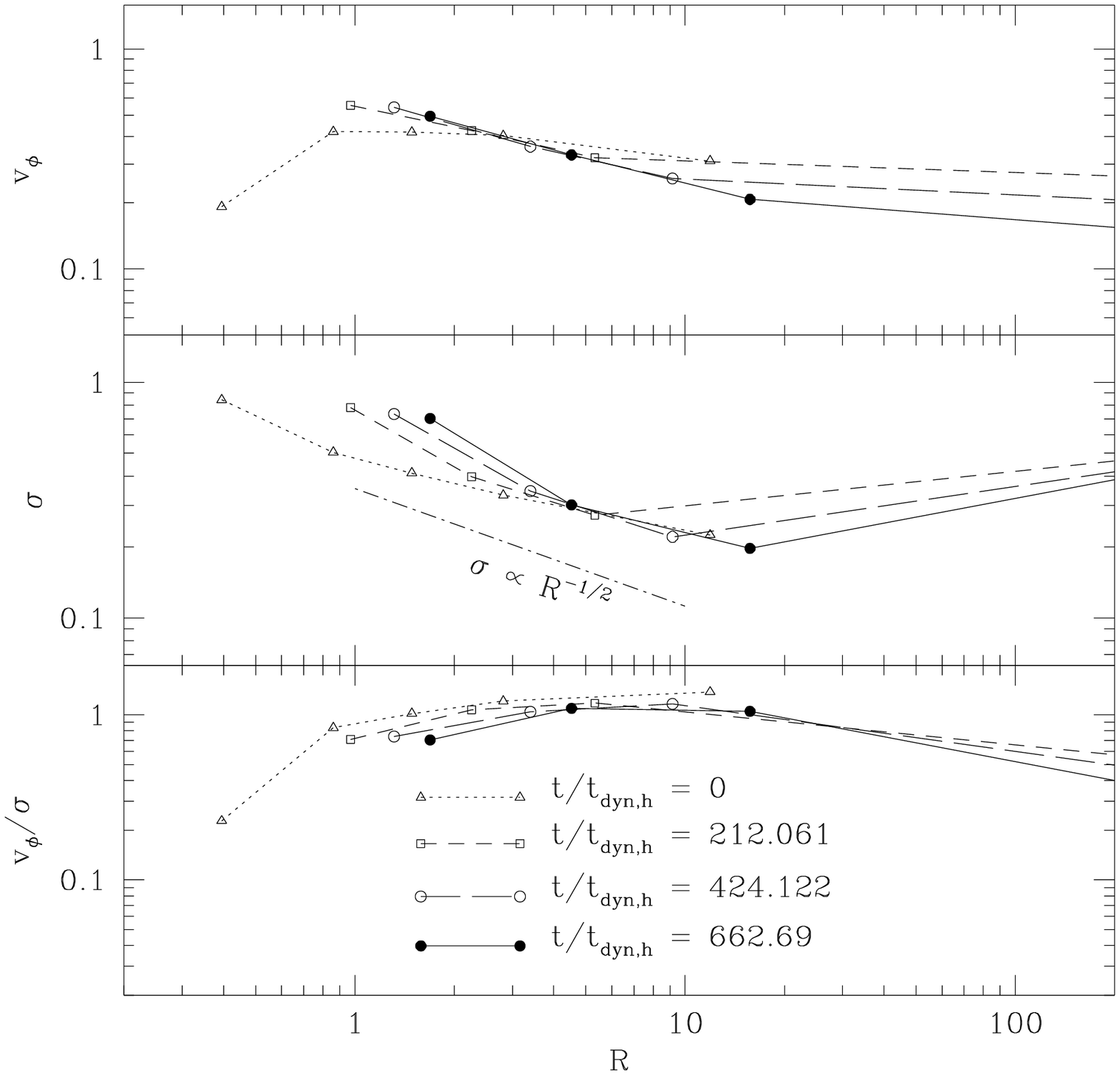}

\figcaption{Rotation-related properties of model XII (equal-mass stars with a
central black hole). \label{vsig}}

\end{document}